\documentclass[preprint,12pt]{elsarticle}

\usepackage{amssymb}
\usepackage{xcolor}
\usepackage{subcaption}
\usepackage{amsthm}
\usepackage{amsmath}
\usepackage{float} 
\usepackage{bm}

\journal{International Journal of Fatigue}

\begin{document}

\begin{frontmatter}



\title{Microstructure sensitive fatigue life prediction model for SLM fabricated Hastelloy-X}


\author[inst1,inst2]{Chandrashekhar M. Pilgar}

\affiliation[inst1]{organization={IMDEA Materials Institute},
            city={ C/Eric Kandel 2},
            postcode={28906}, 
            state={Getafe,Madrid},
            country={Spain}}

\author[inst3]{Ana Fernandez}
\author[inst2,inst1]{Javier Segurado}

\affiliation[inst2]{organization={Universidad Politécnica de Madrid, Department of Materials Science},
            addressline={E.T.S.I. Caminos}, 
            postcode={28040}, 
            state={Madrid},
            country={Spain}}

\affiliation[inst3]{organization={Industria de Turbo Propulsores, Department of Materials and Processes, ITP Aero SAU},
            postcode={48170}, 
            city={Zamudio},
            state={Bizkaia},
            country={Spain}}

\begin{abstract}

A microstructure-sensitive fatigue life prediction framework based on CP-FFT is proposed to study SLM fabricated Hastelloy-X. The microstructure enters in the model through the shape, size and orientation distributions of grains in the RVEs, which are generated from experimental EBSD data. The framework has been applied to specimens built in two different directions with different polycrystalline microstructures and is able to accurately predict their fatigue life, using just two experiments for calibration. The model reproduces the better fatigue performance found experimentally at high stresses for samples built in transverse direction, identifying the origin of this anisotropy in grain aspect ratios.

\end{abstract}



\end{frontmatter}


\section{Introduction}
\label{sec:sample1}
Hastelloy-X is a nickel-based superalloy, primarily strengthened by adding solute elements W, Cr, and Mo \cite{Rowley96,Moataz16} which shows excellent oxidation and corrosion resistance at elevated temperatures, making it suitable for gas turbines and jet engine parts. Moreover, the alloy shows good formability and weldability, making it an ideal candidate for additive manufacturing fabrication, mainly using a selective laser melting (SLM) route \cite{Moataz16}. 

SLM is an additive manufacturing process that can achieve near net shapes of complex geometries with stable properties. In this process, a laser beam follows a predefined 2D path over a platform in a sequential manner. After one traverse of a laser beam, the platform is lowered by the predefined layer thickness. This procedure is repeated till the completion of the final 3D object \cite{Maria19}. SLM fabrication { allows  tailoring of the microstructure by changing parameters such as} laser energy \cite{Maria19}, scanning strategy \cite{Moataz16,Barba19} or build/sample geometries \cite{Moataz16,Barba19}.  The production of net-shape complex components of Hastelloy-X using a SLM process {presents enormous potential for the industry, especially  aerospace, which have already started} to design and fabricate some jet engine parts using SLM fabricated Hastelloy-X. However, strong variability is found  in the monotonic and cyclic/fatigue behavior of {SLM} components depending on the process parameters and part geometry. This variability of properties is mainly attributed to the highly heterogeneous microstructures {resulting from the} complex thermal cycles followed during fabrication  \cite{Lewandowski2016,wilson2017,Pei2019,Kok2018} . The most critical design parameter {for potential high-temperature  applications}  of SLM fabricated Hastelloy-X parts is the fatigue performance. Therefore, models able to predict the fatigue life for {SLM} components accounting for the effect the heterogeneous microstructure are fundamental for the progressive introduction of these components into industrial applications.

\par Multiple authors have studied the cyclic deformation and fatigue behavior of {SLM} components  from an experimental viewpoint 
\cite{aydinoz2016microstructural,kirka2017effect,Kanagarajah2013,Lewandowski2016,Edwards2014,gribbin2016,johnson2017}. Most of these studies reported similar conclusions for a wide range of SLM alloys including IN718 or Ti-6Al-4V \cite{Lewandowski2016,Agius2017}, {namely} an inferior fatigue performance of the {SLM} component compared to wrought and cast ones. In some cases, the reduction of fatigue life can be larger than one order of magnitude  \cite{johnson2017,kelley2016fatigue,Agius2017,Pei2019}.  {This worse fatigue performance of SLM components, particularly in as-built case, has been attributed to the porosity \cite{Li2021}, surface roughness \cite{Konen2016} or residual stresses \cite{Yadollahi2017}, and also to the resulting anisotropy of the part and its orientation {with respect to} the cyclic loading direction \cite{Yadollahi2017,Agius2017}. However, some studies reported the mitigation methodologies for porosity, surface roughness and residual stresses effects on the fatigue lives by optimizing the processing parameters{\cite{Esmaeilizadeh2022}}, heat treatment{\cite{Leuders2013}} and powder composition\cite{Maria19}}.  With respect to the specimen orientation, the fatigue life of vertically {built} specimens has been found  {to be} inferior to horizontally  {built} parts when loaded under stress control \cite{Strner2015,kelley2016fatigue,Edwards2014,Lindstrm2020,Yadollahi2017}. On the contrary, under strain-controlled loading, some studies  \cite{kirka2017effect} show the opposite effect, { vertically built samples having} a superior fatigue performance than horizontally built samples. This fact is justified by the resulting elastic anisotropy, vertically {built} samples have lower elastic modulus and hence underwent lower stress amplitudes than horizontally {built} samples.

In the particular case of AM Hastelloy-X, a few recent studies can be found in the literature which cover its fatigue analysis. Han et al. \cite{Han2018} studied the effect of  {hot isostatic pressing (HIP)} on the fatigue life of AM Hastelloy-X and concluded that the fatigue life could be significantly improved by HIP due to closure of internal pores and relief of residual stresses. Esmaeilizadeh et al.\cite{Esmaeilizadeh2021} studied the fatigue life of AM Hastelloy-X as a function of laser scanning speed at room temperature and concluded that fatigue lives could be improved by optimizing the processing parameters. Wang et al.\cite{Wang2011} highlighted the importance of HIP on fatigue life improvement and reported that the part orientation does not play a significant role in the fatigue life for low applied stress level; however, at higher loads, observed that the fatigue resistance of vertically orientated specimen is inferior to its horizontally oriented counterpart. Lindstrom et al.\cite{Lindstrm2020} reported that the fracture surface of the horizontally oriented specimen had shown more plastic deformation than the vertically oriented specimen.

From the modelling view point, empirical models such as Basquin, Coffin Manson, etc.,  have been used with experimental data to correlate the fatigue life of different additive manufactured alloys \cite{Lee2020,romano2018lcf,Branco2018}, including the study of {SLM} Hastelloy-X by Esmaelizadeh et al.\cite{Esmaeilizadeh2022}.  { However, it is well known that there is a strong influence of the microstructure in the fatigue life of an alloy, and these models cannot account for it because they are empirical and calibrated with experimental data without considering any microstructural aspect.}

{ The influence of the microstructure in the fatigue performance is mainly affecting the nucleation of a crack. The repeated deformation of the polycrystal induces the localization of plasticity at the grain level in persistent slip bands which degenerate finally in microscopic fatigue cracks \cite{McDowell2010}. The regions of the microstructure in which these bands and subsequent cracks are more prone to appear correlate with the points in which some variable related with cyclic plastic deformation is localizing. In order to consider the effect of the microstructures resulting  from SLM fabrication in the fatigue life prediction, micromechanical models which model this mechanisms have to be used.}


In particular, the approaches relying on the use of polycrystalline computational homogenization \cite{Segurado2018} 
to estimate the number of cycles for crack nucleation \cite{McDowell2010} allow to analyze crack nucleation as  {a} function of the microstructure from a statistical viewpoint.
{Under the microstructure sensitive fatigue life approach, fatigue life is predicted by cyclic accumulation of plastic slip and localized stresses that act as a driving force for fatigue crack nucleation on persistent slip bands\cite{lucarini2020upscaling}. The driving force is quantified using Fatigue Indicator Parameters (FIP) that are based on accumulated plastic slip \cite{Manonukul2004,Sweeney2014}, stored strain energy \cite{WAN201490, GUAN201770, CHEN2018213}, dissipated energy \cite{charkaluk2000energetic,COJOCARU20091154,CRUZADO2018b,Cruzado2018} or in general any combination of stress, strain and plastic slip fields \cite{Rovinelli2015}.

The FIP is evaluated at each material point by varying local micro-fields and internal variables, and fatigue life is calculated based on the maximum FIP (hot-spots) in the domain.}
 These models include the relevant microstructural features of the polycrystal under study into Representative Volume Elements (RVE) of the microstructure. The  mechanical response of the RVEs under a macroscopic cyclic history is then simulated using crystal plasticity and some computational homogenization approach to obtain some fatigue indicator parameter that is the driving force for crack nucleation. These micromechanics-based fatigue life prediction models have been successfully applied to a large number of alloys produced by standard fabrication routes, as wrought Ni-based superalloys \cite{Manonukul2004,Castelluccio2014,CRUZADO2018b}. Most of the micromechanical studies of fatigue rely on finite element simulation and only recently Fast Fourier Transform (FFT) solvers have been introduced as an alternative to FEM \cite{Rovinelli2015,Lucarini2018,lucarini2020upscaling}. The use of FFT-homogenization methods  {allow accurate prediction of the micro-field response over many cycles at a fraction of the cost of FE based solvers, making it} possible to simulate larger RVEs and consider more accurate microstructural features \cite{Lucarini2018}.

The application of crystal plasticity models to investigate the effect of the resulting microstructures of AM metals in their fatigue performance is a trending topic nowadays. These models have already been applied to some SLM fabricated  superalloys \cite{PRITHIVIRAJAN2018139,EGHTESAD2021140478,chen2022,GHORBANPOUR2022142913}. In some cases the life prediction is focused on crack propagation \cite{chen2022} stage. Other studies consider crack nucleation but the life prediction model is directly linked with the macroscopic cyclic response \cite{Esmaeilizadeh2021,Esmaeilizadeh2022,Lee2020}, without making use of the distribution of microscopic fields obtained. Only {a} few studies on AM alloys are available which rely on the use of micromechanical fields through fatigue indicator parameters (FIPs) to estimate the fatigue life assuming crack nucleation controlled fatigue. Among these studies, in some cases the real microstructures are not considered and simplified random textures are assumed \cite{cao2022crystal,Ozturk2017}. In other cases, the actual microstructures obtained from EBSD data are included in the model \cite{Ou2020,EGHTESAD2021140478,Prithivirajan2018,Ye2021} but the study is focused only on a particular fatigue regime, either LCF \cite{EGHTESAD2021140478,Ou2020,Ye2021} or HCF \cite{ZHANG2022106577,Han2020}, and a particular loading control, either strain or stress control. 

To summarize, on one hand, there are no available micromechanics based fatigue life prediction approaches available which consider both strain and stress control and a wide range of fatigue lives.  On the other hand, there is still a lack of understanding of the effect that the building direction has on the fatigue performance of SLM alloys, originated in the different microstructures developed during SLM fabrication.

In this work, we present a micromechanics based fatigue life prediction approach valid for stress and strain control and a wide range of fatigue lives. From a numerical viewpoint, FFT based polycrystalline homogenization is used to simulate the cyclic response of the specimen. This framework is then applied to analyze the effect of the SLM resulting microstructures, including grain aspect ratios and orientation distributions. The model is applied to study the fatigue life of Hastelloy-X specimens built in different directions \cite{Pilgar2022} including its validation using experimental data under strain and stress control at 750 $^{\circ}$C for a wide range of fatigue cycles from very low LCF to more than $10^5$ cycles.

\section{Experimental study}
\subsection{Material and fabrication}
This work uses a selective laser melting(SLM) route  to manufacture the Hastelloy-X samples, whose nominal composition is given in \cite{Maria19}.

\begin{figure}
	\centering
		\includegraphics[width=\textwidth]{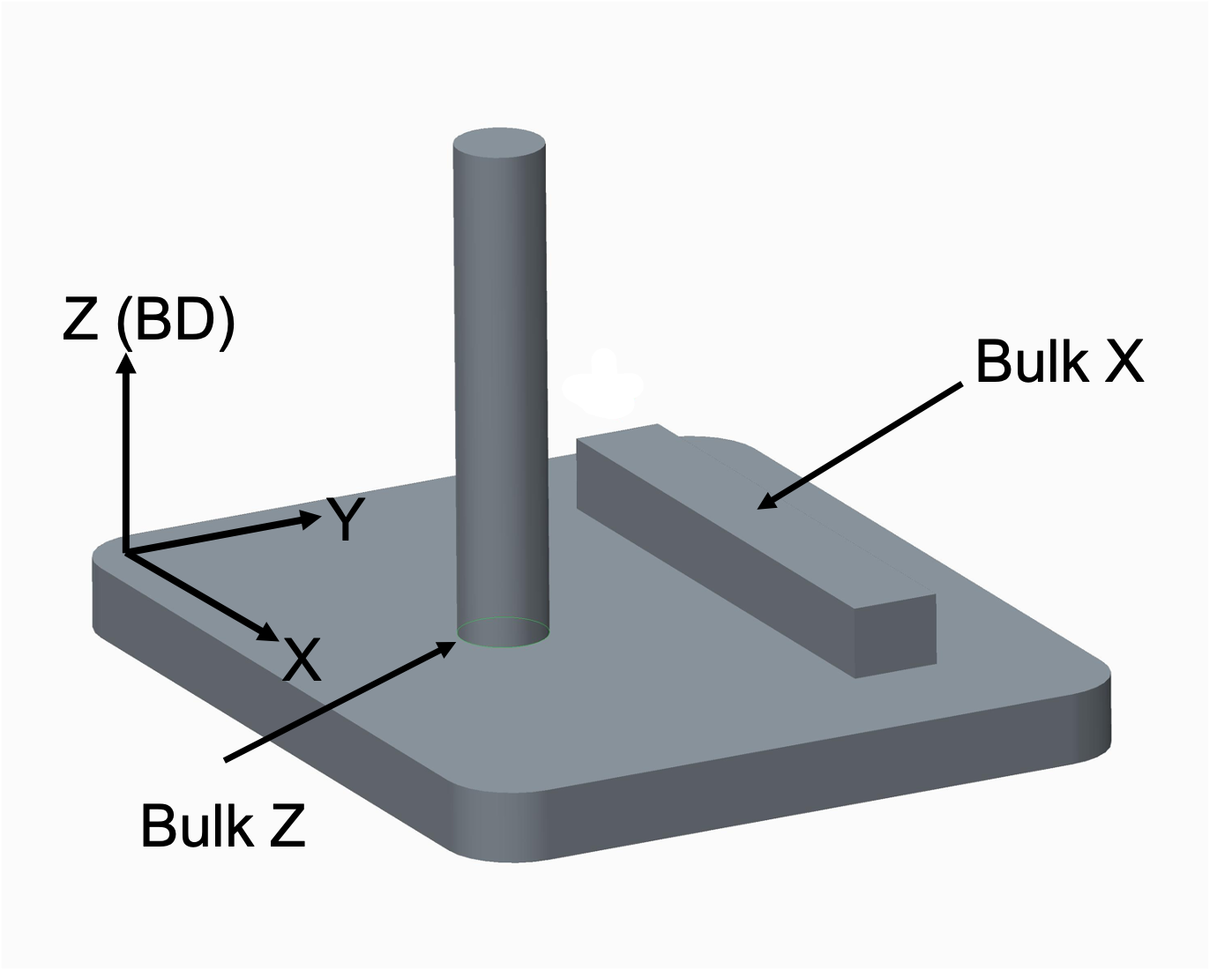}
	\caption{\em{The SLM platform shows different orientations of the bulk samples fabricated using the SLM process along with the build direction.}}
	\label{fig:Samples}
\end{figure}

The bulk samples were built layerwise in the Z direction using the Renishaw RenAM 500Q machine, keeping a constant layer thickness of 60 $\mu m$ and a rotation of the laser scanning direction  between consecutive layers of 67$^{\circ}$   {(see Fig. \ref{fig:Samples})}.  Hastelloy-X powder was made via the Ar gas atomization process, providing a powder size in the range of { $15 \; \mu m $ to $53 \; \mu m$}. The parameters of the SLM process, such as laser power, scan speed, hatch distance, and volumetric energy density, were well optimized to achieve the maximum part density. The different fatigue specimens were fabricated through the SLM process by ITP Aero in the shape of bulk blocks.
The bulk samples were fabricated in two different orientations {(see Fig.\ref{fig:Samples})}:
\begin{itemize}
    \item Bulk Z specimen is built in a cylindrical shape, oriented parallel to the build direction. 
    
    \item {Bulk X specimen is printed with a rectangular shape with dimensions 
    {15 mm$\times$15 mm$\times$100 mm}, oriented perpendicular to the build direction.} 
\end{itemize} 

\begin{figure}
	\centering
	\includegraphics[width=\textwidth]{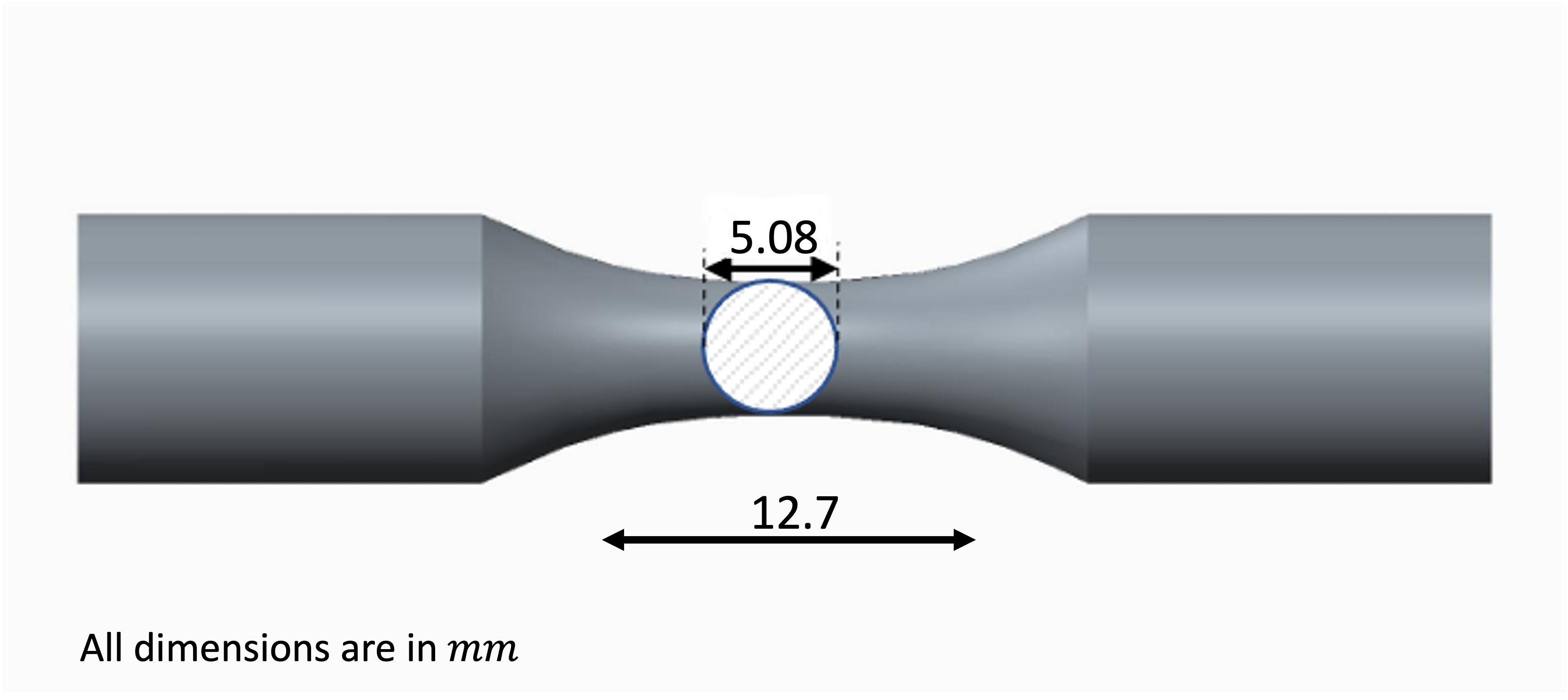}
	\caption{\em{ The cylindrical dog-bone-shaped specimen, used for the uniaxial tensile fatigue test.}} 
\label{fig:fatigue_sample}
\end{figure}

Coupons necessary for fatigue tests were machined from above-built bulk samples in cylindrical dog-bone shapes with the same gauge length ({12.7 mm}) and diameter ({5.08 mm}) shown in Fig.\ref{fig:fatigue_sample}. Finally, the samples were heat treated at 1170$^{\circ}$C / 30 minutes and followed by gas cooling. {
The SLM thermal history was removed by this treatment and the SLM process dendrites were dissolved. This annealing also significantly minimises the residual stresses of Hastelloy-X, as observed in Karapuzha et al. \cite{ShajiKarapuzha2022}.  { It is also interesting to note that precipitates were found in significant quantities in the Ni-based as-build samples, but after the heat treatment the presence of precipitates in the alloy was strongly reduced or even suppressed.}

}

\subsection{Microstructure}
{ The resulting porosity was obtained using a microCT analysis and was below 1\% in all the samples.} With respect to the polycrystalline microstructure, it is well known that changes in the printing process parameters generate microstructures with different grain elongation and textures. The bulk samples showed different grain structures in BD than in the XY plane. The elongated grains are observed in the build direction. The direction of epitaxial grain growth depends on the direction of maximum heat flow. The maximum heat flow in the SLM process is observed in the building direction. However, the equiaxed grains are observed in the lateral direction.  


The grain size, shape, and orientation distributions were obtained using Electron Backscatter Diffraction(EBSD) imaging technique.
Concerning the grain morphology, the Bulk Z sample reveals elongated grain arrangement in the build direction along with an equiaxed grain setup in the XY section (see Fig. \ref{fig:EBSD_exp}). The Bulk X specimen exhibits more equiaxed grain arrangement in both sections. The average grain aspect ratio was 1/1/2 and 1/1/1.5 in X-Y-Z axis, for Bulk-Z and Bulk-X samples respectively.

The diameter of the equiaxed grains in the XY plane follows the log-normal distribution. The log-normal distribution parameters(mean and standard deviation) for both bulk samples are enlisted in Table \ref{tab:lognormal}. The surface roughness is always present in the {SLM} samples due to partially melted and unmelted powder on the surface. Therefore, the samples are machined from all directions to minimize the surface roughness impact on the fatigue test.


\begin{table}[H]
	\centering
	\caption{The lognormal distribution of { grain diameter} of the bulk samples}
	\vspace{1mm} 
	\label{tab:lognormal}
	\begin{tabular}   {lll}
	\hline
		Sample  & Bulk Z & Bulk X  \\
		Average($\mu m$)  & 17.94 & 20.13  \\
		Standard deviation & 0.94 & 0.76  \\
		\hline
		
	    \hline
	\end{tabular}
\end{table}

\begin{figure}[htbp]
	\centering
	\includegraphics[width=\textwidth]{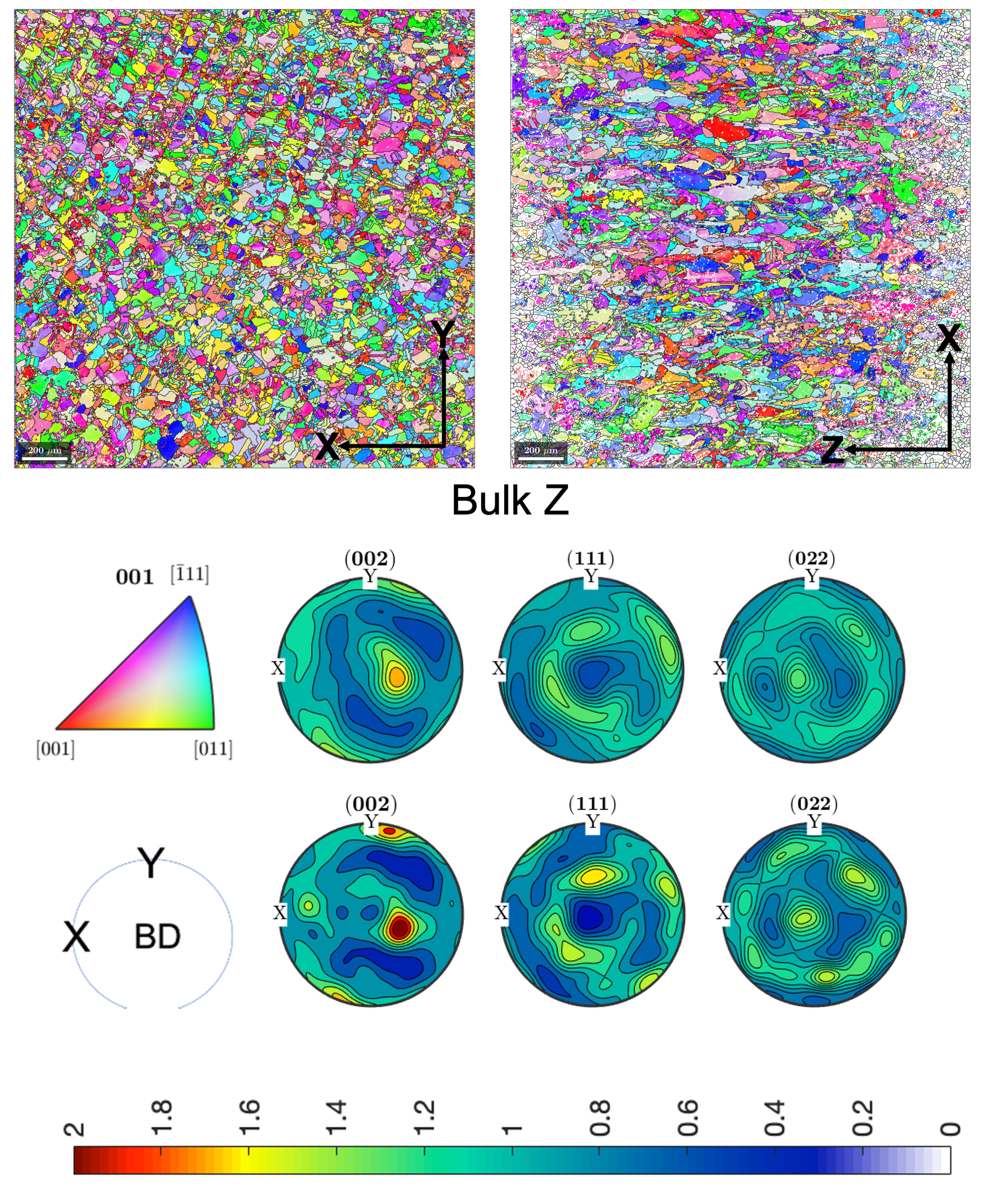}
\caption{\em{ EBSD images of Bulk Z (top)  specimen in XY and XZ sections respectively. The colors of the maps suggest the orientation of the building direction compared to the reference frames following the IPF triangle. Pole figures using all the orientations (second) and a reduced set obtained after sampling the ODF(bottom), for Bulk Z specimen.}} 
\label{fig:EBSD_exp1}
\end{figure}

\begin{figure}[htbp]
	\centering

	\includegraphics[width=\textwidth]{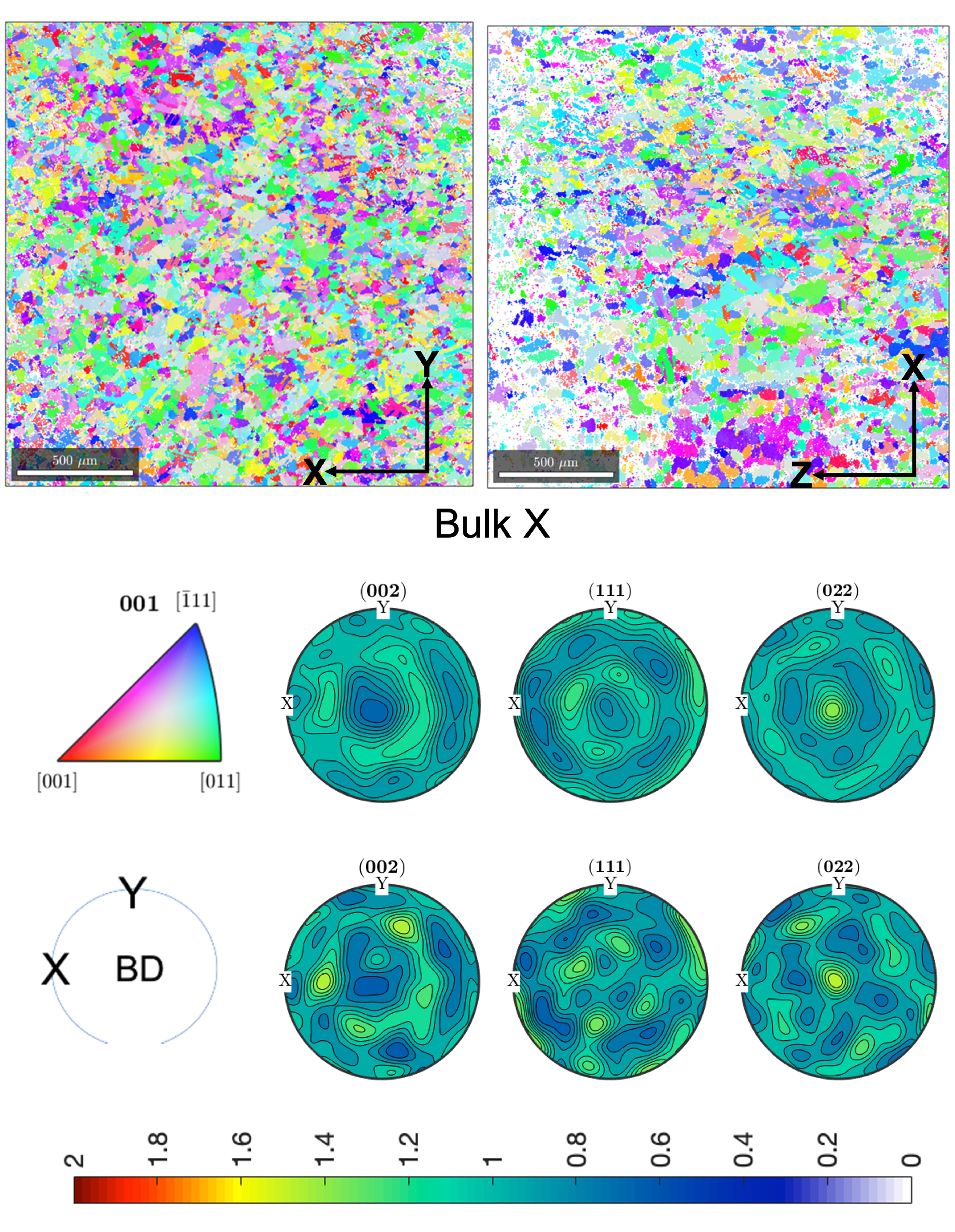}

\caption{\em{ EBSD images of Bulk X (top) specimen in XY and XZ sections respectively. The colors of the maps suggest the orientation of the building direction compared to the reference frames following the IPF triangle. Pole figures using all the orientations (second) and a reduced set obtained after sampling the ODF(bottom), for Bulk X specimen.}} 
\label{fig:EBSD_exp}
\end{figure}

To obtain the orientation distribution of each grain in the microstructure, the EBSD analysis was performed on the samples before the fatigue testing. For this purpose, a SEM with an EBSD detector was used. A confidence index (CI) criteria was used to ensure the accuracy of the measured Euler angles. The Euler angles are measured only if the CI at a pixel is greater than 0.02. 
The OIM analysis software generates the resultant EBSD orientation maps and M-TEX \cite{bachmann2011grain} generates corresponding pole figures in Z-direction, which are depicted in {Fig.\ref{fig:EBSD_exp1} and \ref{fig:EBSD_exp}  for the XY and XZ sections of Bulk Z and Bulk X samples respectively}. The colors in the EBSD maps that follow the IPF triangle gives the orientation of the building direction concerning the crystal reference frame, and the colors in the pole figures follow the color bar shown at the bottom of {Fig.\ref{fig:EBSD_exp1} and  \ref{fig:EBSD_exp}}. The pole figures depict that both bulk samples have a random texture along the XY and XZ sections.

\subsection{Mechanical Characterization}
The uniaxial cyclic performance of SLM fabricated Hastelloy-X at 750$^\circ$ was experimentally determined by cyclic tests performed in a wide range of strain  and stress amplitudes. The tests were carried out according to the standards PrEN3874-98 and ASTM E606. The results of the fatigue testing campaign will be presented in the result section.

\section{Micromechanics Based Fatigue Life Prediction Model}
A computational polycrystalline homogenization framework based on the {crystal} plasticity and FFT homogenization (CP-FFT)  is used to simulate the cyclic response for SLM Hastelloy-X under strain and stress-controlled loading. The FFT-based homogenization code FFTMAD \cite{Lucarini2019} is used to simulate the cyclic response of Representative Volume Elements (RVE) of the microstructures considered.

The synthetic RVEs contain grain size distribution, grain aspect ratio, and orientations representative of the actual SLM microstructures experimentally characterized. This section will describe the RVE generation, crystal plasticity model that accounts for the cyclic deformation of grains, the FFT homogenization framework, and the quantification methodology to predict fatigue life based on the FIP. 

\subsection{RVE Generation}
The digital representation of RVE corresponding to the SLM Hastelloy-X microstructure is done to statistically include the grain size, shape, and orientations obtained from the EBSD. The detailed RVE generation methodology is enclosed in our previous work \cite{Pilgar2022}, and briefly recalled here for completeness. 

 The 2D grain size distribution is experimentally obtained from the XY section of EBSD images (see Fig 4) and is first converted to the apparent 3D grain size, assuming spherical grains, using the software \emph{GrainSizeTools}. \cite{lopez2018grainsizetools}. The 3D distributions are then used as input to generate digital RVEs consisting of equiaxial polycrystals. To this aim, a cloud of points and their corresponding weights are generated to define a weighted Voronoi tessellation which reproduced the targetgrain size distribution \cite{Lucarini2019}.
 
\begin{figure}
	\centering
		\includegraphics[width=\textwidth]{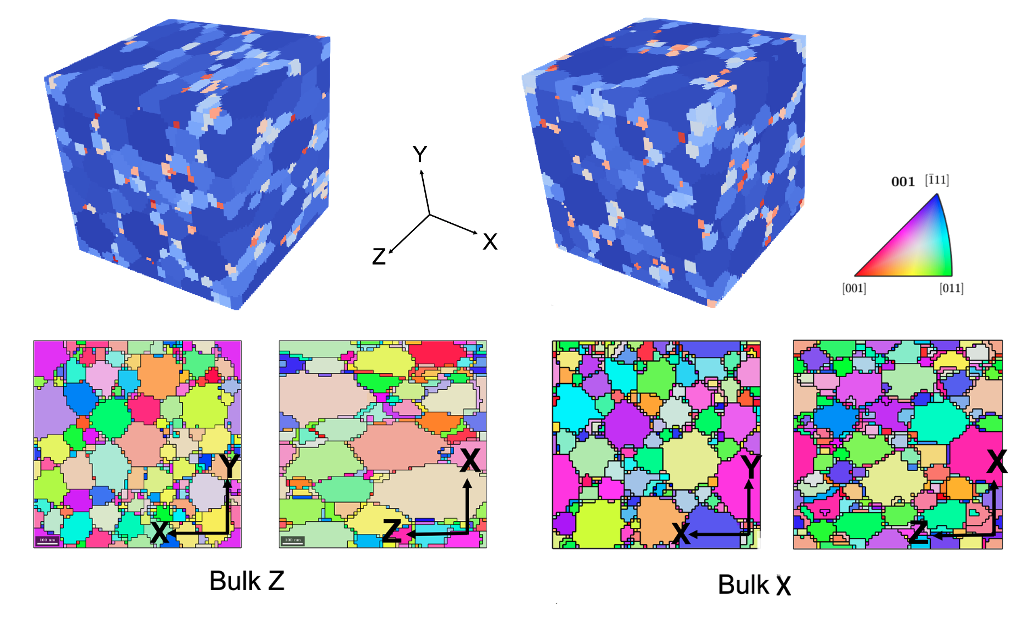}
        \caption{\em{ 3D images of RVEs of polycrystalline SLM Hastelloy-X for Bulk Z (top left) and Bulk X (top right). Note that colors indicate the grain numbers. EBSD maps of Bulk Z(bottom left) and Bulk X (bottom right) samples in XY and XZ sections.The colors of the maps suggest the orientation of the building direction compared to the reference frames following the IPF triangle.  }} 
\label{fig:RVE_pred}
\end{figure}

 Finally, RVEs are scaled proportional to the aspect ratios measured. The resulting synthetic RVEs are shown in Fig.\ref{fig:RVE_pred} with XY and XZ sections.

\subsection{Crystal Plasticity Model} \label{crystal_plasticity}
The crystal plasticity (CP) model developed by Cruzado et al.\cite{Cruzado2017} for the cyclic response of Inconel 718 is used in this study to simulate the elasto-viscoplastic cyclic behavior of Hastelloy-X crystals. This model allows to reproduce the isotropic and kinematic hardening and mean stress relaxation effects which might appear under cyclic loading. { The model selected is size independent because the differences in grain sizes are below 10\%, which would have a negligible influence in the fatigue life. Moreover, kinematic hardening is considered by a flow rule, in contrast to some strain gradient models used in the context of fatigue \cite{WAN201490} in which kinematic hardening is obtained as a result of the computing explicitly the GNDs. The reason for this simplification is having a computationally efficient model which allow us to fulfil the two main requirements of our study  (1) the use of realistic,  statistically representative and well discretized microstructures and (2) considering fatigue life prediction in a statistical manner.}

\par A multiplicative decomposition of the deformation gradient $\textbf{F}$ into its elastic ($\text{\textbf{F}}^e$) and plastic ($\text{\textbf{F}}^p$) components is assumed, 
\begin{equation} \label{eq1}
    \textbf{F}=\text{\textbf{F}}^e \text{\textbf{F}}^p
\end{equation}


The plastic velocity gradient $\textbf{L}^p$ is computed in the intermediate (relaxed) configuration as function of the shear rates ${\dot{\gamma}}^{\alpha}$ on all the slip systems $\alpha$,
\begin{equation} \label{eq2}
    \textbf{L}^p=  \dot{\textbf{{F}}}^p \textbf{F}^{p^{-1}}=
   \sum_{\alpha} \dot{\gamma}^{\alpha} \text{\textbf{s}}^\alpha \otimes \text{\textbf{m}}^\alpha
\end{equation}\\
where, $\text{\textbf{s}}^\alpha$ and $\text{\textbf{m}}^\alpha$ are the slip and slip normal directions, respectively, in the initial configuration.

The second Piola-Kirchoff stress $\textbf{S}$ depends linearly on the Green-Lagrange strain $\textbf{E}^e$ through the elastic stiffness tensor $\mathbb{C}$, which is given by
\begin{equation} \label{eq6}
    \textbf{S}= \mathbb{C} \textbf{E}^e  
\end{equation}
and the Cauchy stress tensor is calculated as,
\begin{equation} \label{eq8}
  \sigma =  \frac{1}{ {J}^e } \textbf{F}^e    \textbf{S} \textbf{F}^{e^{T}}
\end{equation}
, being ${J}^e$ the determinant of $\textbf{F}^e$.

A power-law is used to define the slip rate in each slip system as, 
\begin{equation} \label{eq9}
 \dot{\gamma}^{\alpha}= \dot{\gamma}_{0} \left( \frac{|\tau^{\alpha} - \chi^{\alpha}|}{ {g}^\alpha } \right)^{\frac{1} {m}} \text{sign} (\tau^{\alpha}-\chi^{\alpha})
 \end{equation}
where $\dot{\gamma}_{0}$ and $m$ are, respectively, the reference strain rate and the strain rate sensitivity parameters. $\tau^{\alpha}$ is the resolved shear stress, defined as,
\begin{equation} \label{eq7}
  \tau^{\alpha} = \textbf{S} :  (  \text{\textbf{s}}^\alpha \otimes \text{\textbf{m}}^\alpha).
\end{equation}
The functions ${g}^{\alpha}$ and ${\chi}^{\alpha}$ are the critical resolved shear stress and backstress on the $\alpha$ slip system respectively.

The evolution of the backstress on each slip system defines the kinematic hardening. In the present model \cite{Cruzado2017}, a simplified version of the Ohno Wang Model \cite{Ohno1993} limited to the first two terms is used,
\begin{equation} \label{eq10}
 \dot{\chi}^{\alpha}= c \dot{\gamma}^\alpha - d\chi^\alpha |\dot{\gamma}^\alpha| \left( \frac{|\chi^\alpha|}{c/d} \right)^k
 \end{equation}
where $c,d,$ and $k$ are material parameter. $k$ controls the mean stress relaxation, while $c$ and $d$ stand for direct hardening and dynamic recovery modulus, respectively. 

The evolution of the Critical Resolved Shear Stress (CRSS) for a given slip system, $g^{\alpha}$, defines the isotropic hardening and is given by
\begin{equation} \label{eq11}
  \dot{g}^{\alpha}=\sum_{\beta} q_{\alpha \beta} h(\gamma_a) |\dot{\gamma}^{\beta}|
\end{equation}
where $q_{\alpha \beta}$ are the latent hardening coefficients and $h$ is the self hardening modulus which follows the Voce hardening model \cite{Tome84}, given by
\begin{equation} \label{eq13}
   h({\gamma_a})= h_s+ \left[h_0-h_s+\frac{h_0 h_s \gamma_a}{\tau_s -\tau_0}\right]exp^{\left( \frac{-h_0 \gamma_a}{\tau_s-\tau_0}  \right)}
\end{equation}
where, $\tau_0,\tau_s,h_0$ and $h_s$ are the hardening parameters and the accumulated plastic shear $\gamma_a$ is computed as,  
\begin{equation} \label{gamma_tot}
\gamma_a=\sum_\alpha \int_{0}^{t} |\dot{\gamma}^\alpha| \mathrm{d}t.
\end{equation}

\subsection{FFT based computational homogenization framework}

The FFT homogenization code, FFTMAD \cite{Lucarini2018,Lucarini2019} is used to perform the virtual fatigue test that provide the homogenized performance of polycrystals and the localized stress and strain fields. 

The polycrystal response is obtained by simulating the  RVEs generated, which are discretized into a regular array of voxels. The constitutive model summarized in the previous section enters through a \emph{UMAT} subroutine \cite{Cruzado2017}. The FFT homogenization problem in finite strains is solved using the Galerkin FFT method \cite{Vondejc2014,Zeman2017,Geus2017}. The differential operators in the FFT solver used a finite differentiation approach using the rotated scheme developed in  \cite{Willot2014}. Strain, stress and mixed control macroscopic boundary conditions are introduced following the technique proposed by Lucarini et al. \cite{Lucarini2018}. According to this technique, the external macroscopic loading is controlled in terms of macroscopic deformation gradient $\overline{F}_{ij}(t)$ or first Piola-Kirchoff stress components $\overline{P}_{IJ}(t)$, for strain and stress-controlled fatigue respectively \cite{Lucarini2019}. The loading process is discretized in increments and the resulting non-linear system of equations for each time increment defines the deformation gradient which provides stress equilibrium. The non-linear equation is solved using Newton Raphson  \cite{Geus2017} and the conjugate gradient method is used as linear solver for each Newton iteration. 

\subsection{Fatigue Indicator Parameters and Life Prediction}
{ The successive deformation of the polycrystal induces the localization of plasticity in persistent slip bands which degenerate finally in microscopic fatigue cracks \cite{McDowell2010}. The areas of the microstructure in which these bands and subsequent cracks will appear correlate with regions where some Fatigue Indicator Parameter (FIP) is maximum. Moreover, quantifying the value of this FIP in that hot-spots allows to estimate the number of cycles to crack nucleation.}
 { 
In the current study, the Fatigue Indicator Parameter through the microstructure is obtained from computation homogenization of a full cycle. The FIP distribution within the RVE was computed from the variation of micro-fields and the state variables at every material point. This distribution defines the hot-spots in which fatigue cracks are more likely to appear and the fatigue life is predicted based on the most critical point. The most common FIPs found in the literature are, accumulated plastic strain per cycle \cite{Manonukul2004,McDowell2010,Sweeney2014}, the plastic stored energy per cycle \cite{WAN201490,CHEN2018213}, accumulated internal strain energy and
some other parameters. }

{
In this work, plastic work per cycle $W_{cyc}(\textbf{x})$ is chosen as local FIP, as it has shown to correlate well the fatigue life  in many different materials \cite{charkaluk2000energetic,COJOCARU20091154,CRUZADO2018b,Cruzado2018}. This election is merely practical and other FIP would probably lead to similar results. As an example, the stored energy criterion \cite{WAN201490,CHEN2018213}, would  provide very close number of cycles, since the stored energy is a small fraction of the total plastic work per cycle.}

This FIP is defined as
\begin{equation} \label{eq14}
W_{cyc}(\textbf{x})= 
\int_{cyc} \tau_{\alpha}(\textbf{x}) \dot{\gamma}_{\alpha}(\textbf{x}) dt 
\end{equation} 
 at each point $\textbf{x}$ in the RVE being $\tau_\alpha$ and $\dot{\gamma}_{\alpha}$ the resolved shear stress and shear strain rate respectively. 
 Eq.\ref{eq14} is applied at the center of each voxel to obtain a local map of FIP values in the RVE. 

 However, the crack initiation on the persistent slip bands is a non-local phenomenon, as shown for example in Castelluccio et al. \cite{Castelluccio2014}. For this reason, and considering also the strong mesh dependence of the local FIP maps, non-local FIP maps are defined as the volume averaged of local FIPs over the grains or bands within a grain parallel to the slip planes with a constant thickness \cite{Castelluccio2014,CRUZADO2018b}. { This type of non-local variable corresponds to an  non-local integral type model. Alternative non-local FIP by means of gradient approaches can be derived, as proposed  in \cite{XU2021102970}}

 The estimation of the number of cycles for crack nucleation is done based on the most critical point of an RVE. In this work, the maximum of the non-local FIPs defined at each band in the RVE is taken as driving force. More details on the geometrical definition in bands can be found in \cite{CRUZADO2018b}. Each grain is divided in four families of bands, being $nb$ the total number of bands in the RVE. The FIP of the full RVE is then defined as 
  \begin{equation} \label{eq15}
W_{cyc}^b= \max_{i=1,nb} \left\{   \max_{\beta_{i}} \frac{1}{V_i}
\int_{V_i} W_{cyc}^{\beta_i}(\mathbf{x})dV_i\right \}
\end{equation} 
where $1\leq i \leq nb$ refers to each of the bands in the RVE, $\beta_i= 1, 2, 3$, corresponds to the three different slips systems contained in the slip plane parallel to that band and $V_i$ is the volume of the band. 

Normally, crack nucleation is directly related to the cyclic FIP obtained from Eq.\ref{eq15} using some empirical relations. However, in this work, a different methodology has been used to predict both fatigue life for stress- and strain-controlled loading, keeping a similar definition of FIP. 

 \subsubsection*{Strain Controlled Loading}
 The value of the FIP in a stable cycle can be used to estimate, based on extrapolation, the number of cycles for crack nucleation. The relation between this value and the number of cycles for nucleation can be based on a linear extrapolation \cite{Manonukul2004} 
 or on other relation, as the power law in \cite{CRUZADO2018b,Cruzado2018}. In the latter case, the value of the FIP in a stable cycle, $W_{cyc}$, is linked to the number of cycles for nucleation $N_i$ as
\begin{equation} \label{eq16}
N_i= \frac{W_{crit}}{{(W_{cyc})}^m}, 
\end{equation} 
where $W_{crit}$ and $m$ are two parameters of the material. This biparametric fatigue life prediction law requires the use of two independent fatigue experiments \cite{CRUZADO2018b} to be calibrated.

The use of the previous approach implies direct simulation of the RVE deformation until a stable hysteresis cycle is reached. Under strain control, hysteresis evolves from cycle to cycle, and stabilization can be reached after a few cycles in some cases to thousands of cycles in other cases, depending on the material and loading conditions \cite{Manonukul2004}. To limit the computational cost of these simulations, different strategies have been developed in the context of polycrystalline homogenization, such as the wavelet transformation-based multi-time scaling algorithm (WATMUS) \cite{Joseph2010} or the linear extrapolation of internal variables or cycle jumps \cite{Cruzado2017}. 

\subsubsection*{Stress Controlled Loading} \label{stress_controlled}
Contrary to strain-controlled tests, under stress-control loading, the plastic flow is concentrated mainly in the first few cycles. From a computational point of view, this usually results in a relatively fast transition to an approximate elastic cyclic macroscopic response. Although microplasticity can still be observed after the shakedown regime, the associated FIPs are three to four orders of magnitude lower than the FIPs in plastic regimes, too small for an accurate prediction of crack nucleation based on the extrapolation approaches used for strain control.

As an alternative, we consider that the total accumulated plastic deformation (also related to the stored energy) until the shakedown limit is the precursor to the nucleation of a crack. Therefore, in this work, the total accumulated plastic deformation, $W^{acc}_{p}$, is proposed as FIP to correlate the number of cycles for crack nucleation and the microplasticity effects after shaking are neglected in this correlation.  

The approach proposed here first introduces a criterion to quantify, based on microscopic fields in an RVE simulation, the number of cycles $n=ns$ at which elastic shakedown occurs.
 \begin{equation} \label{eq18}
ns \ \text{such} \ \text{that} \ W_{cyc}(n=ns)=   10^{-4} W_{cyc}(n=1).
\end{equation} 
where $W_{cyc}(n)$ is the FIP value in the RVE (Eq. \ref{eq15} 
) on the $n^{th}$ cycle. Then, the accumulated plastic deformation, $W^{acc}_{p}$, is given as
 \begin{equation} \label{eq17}
W^{acc}=  \sum_{n=1}^{n=ns} W_{cyc}(n)
\end{equation} 

A power-law function is proposed (Eq. \ref{eq19}) which also includes two fitting parameters, $W^{crit}_{\sigma}$, and $nk$  to link the accumulated plastic FIP, $W^{acc}_{p}$, and the fatigue crack initiation life $N_i$. 
 \begin{equation} \label{eq19}
N_i= \frac{W^{crit}_{\sigma}}{{(W^{acc}_{p})}^{nk}}
\end{equation} 

\section{Results}
\subsection{Experimental results}
Uniaxial tensile fatigue tests were performed using strain control in the case of  Bulk X samples and under stress control for both Bulk X and Bulk Z. The cylindrical-shaped bulk sample (Bulk Z) was subjected to uniaxial tensile cyclic loading in the building direction (Z), while the rectangular-shaped sample ( Bulk X) was subjected to uniaxial tensile cyclic loading in the X-direction. In the case of strain controlled tests, the hysteresis stress-strain cycle was registered cycle by cycle in order to obtain the cyclic plastic response. From now on, due to the confidentiality agreement signed with the industrial partner, all data are normalized by constant normalization factors. The normalization factor $\Delta \epsilon_{min}$ corresponds to the minimum applied strain range under strain loading conditions. For stress, the normalization factor is the critical resolved shear stress, $\tau_0$, identified for Hastelloy-X grains at $750^{\circ}$C.

\begin{figure}
\begin{subfigure}{.5\textwidth}
  \centering
  \includegraphics[width=1\linewidth]{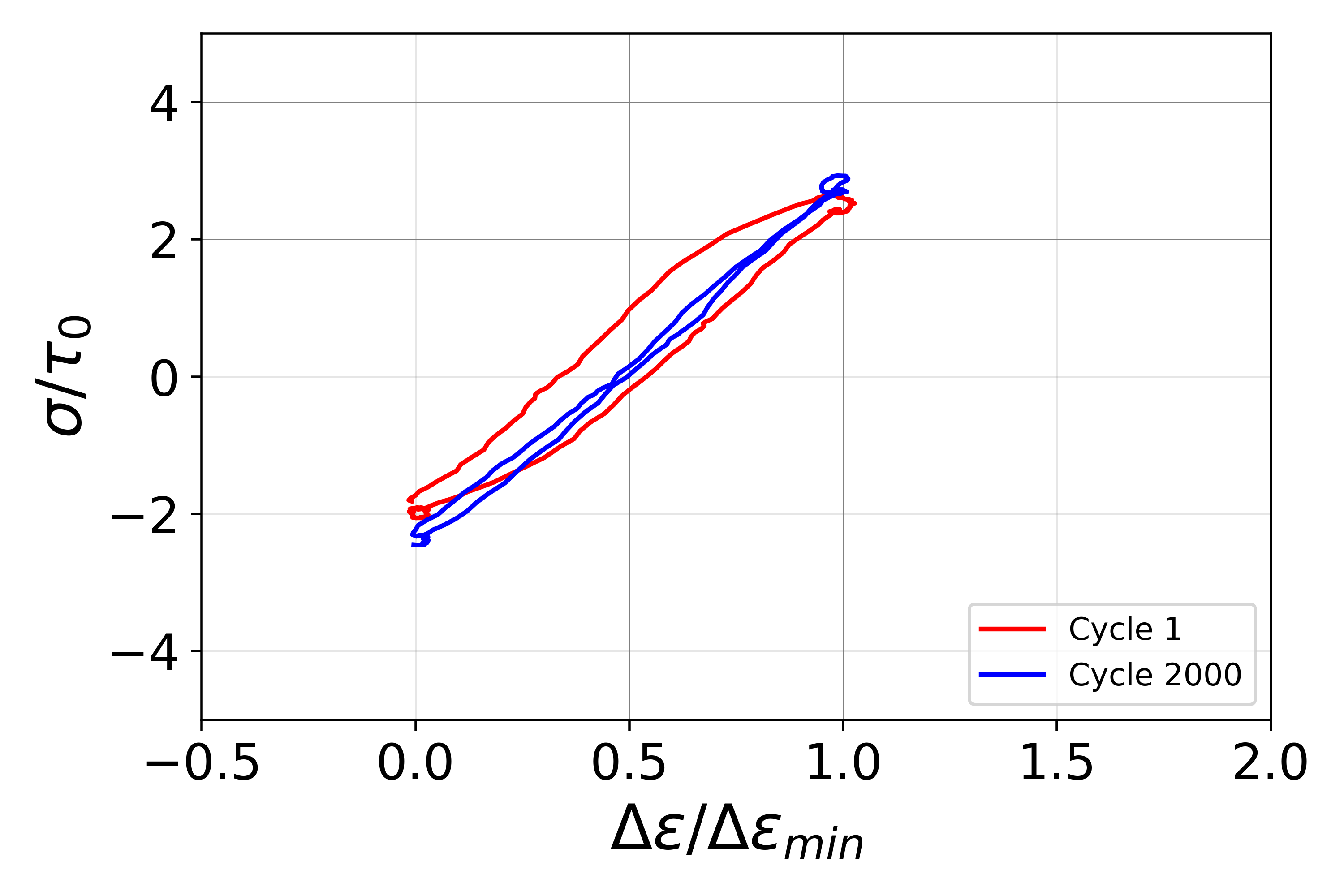}
  \caption{}
  \label{}
\end{subfigure}%
\begin{subfigure}{.5\textwidth}
  \centering
  \includegraphics[width=\linewidth]{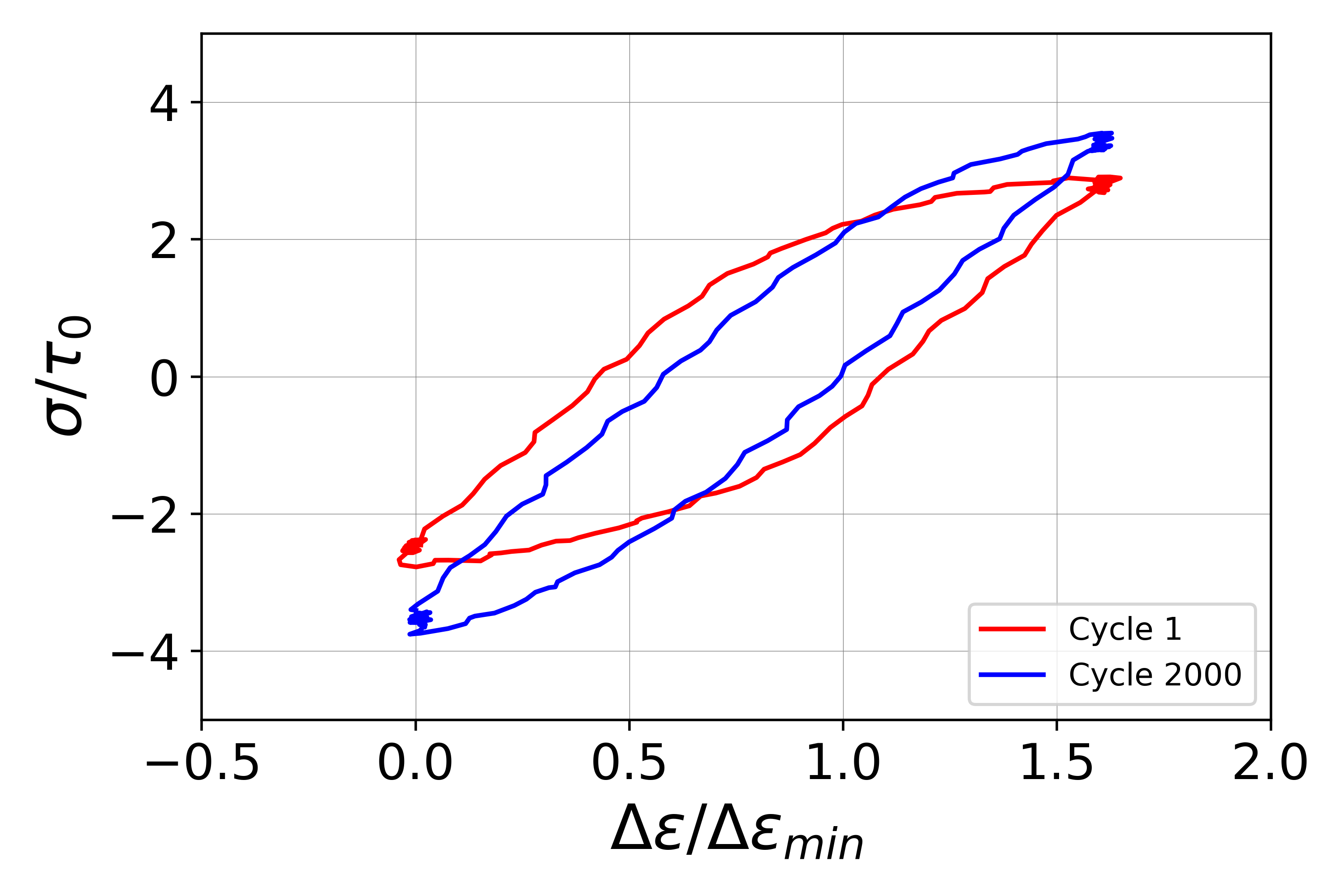}
  \caption{}
  \label{}
\end{subfigure}
\caption{\emph{ Under strain controlled loading, the stress-strain cycle for the first(shown in red color), and $N=2000$ cycles(shown in blue color) are represented for two strain ranges, (a) ${\Delta \epsilon}/{\Delta \epsilon_{min}}$= 1, and (b) ${\Delta \epsilon}/{\Delta \epsilon_{min}}$= 1.6 }}
\label{fig:Comp_Hyst_strain}
\end{figure}

The cyclic uniaxial tests under strain control were carried out with strain amplitudes, $\frac{\Delta \epsilon}{\Delta \epsilon_{min}}$= 1, 1.1, 1.2, 1.3, 1.4, 1.5, 1.6, 1.7 and, 1.8 under strain ratio $R_\epsilon$=0. The hysteresis cycles show a combination of isotropic and kinematic hardening and finally collapse in a stable cycle. An example of these experimental results is represented in Fig. \ref{fig:Comp_Hyst_strain}, where the stress-strain cycle for the first and $N=2000$ cycles (when the cycle became stable) is represented for two strain ranges, $\frac{\Delta \epsilon}{\Delta \epsilon_{min}}$= 1 (Fig.\ref{fig:Comp_Hyst_strain}(a) and $\frac{\Delta \epsilon}{\Delta \epsilon_{min}}$= 1.6 (Fig.\ref{fig:Comp_Hyst_strain}(b)). The rest of the results will be represented together with the model predictions in the next section. 

In the case of stress-controlled tests, the applied uniaxial stress range considered was  $\frac{\Delta \sigma}{\tau_0}$= 3.2, 3.57, 3.93, 4.29, 4.64, and 5 under stress ratio $R_\sigma$=0.03.  
{ Both stress and strain controlled tests were carried out at 750°C using a trapezoidal (1s-1s-1s-1s) waveform with a frequency of 0.25Hz and a stress concentration factor ($K_t=1$).}

\begin{figure}
\begin{subfigure}{.5\textwidth}
  \centering
  \includegraphics[width=1\linewidth]{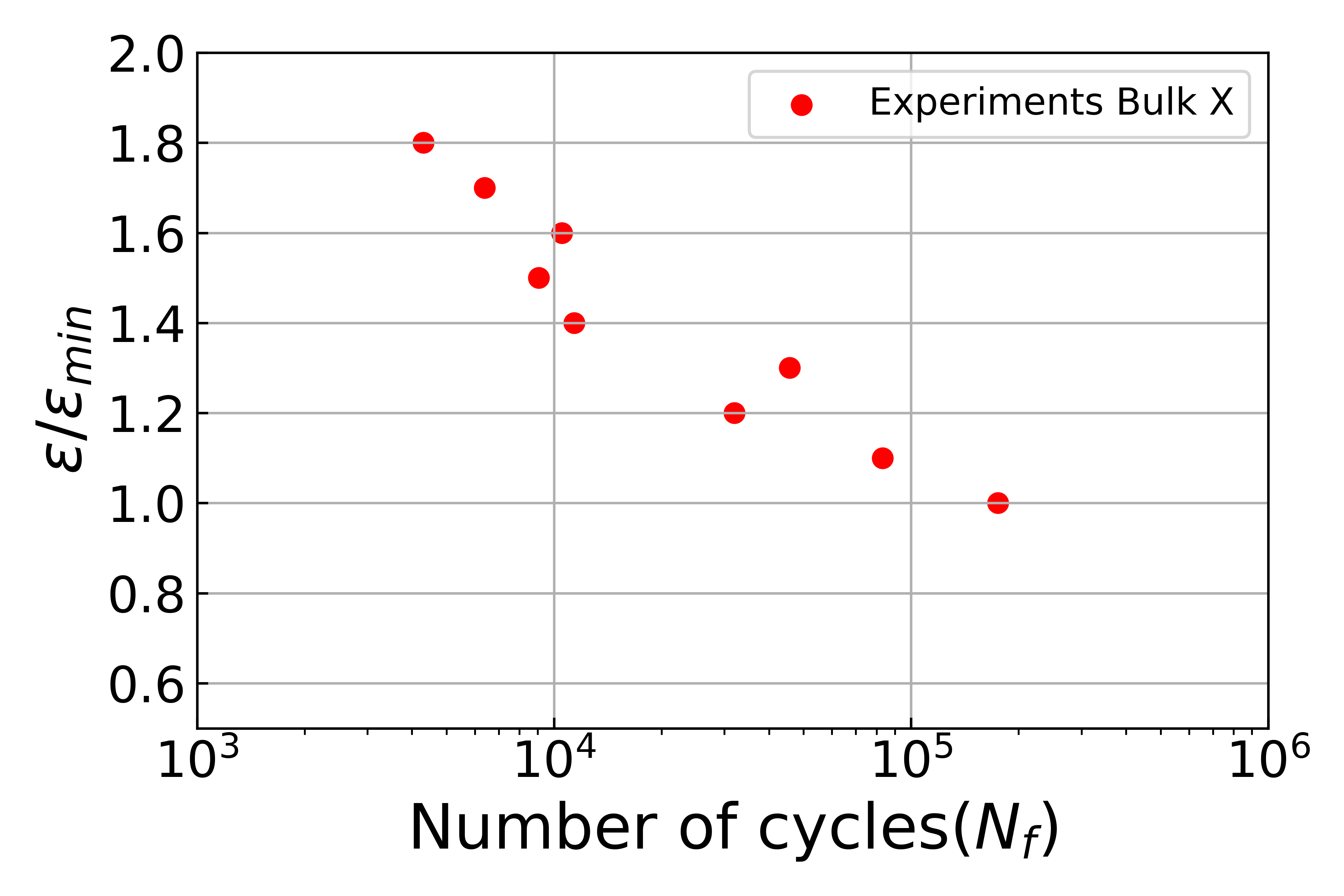}
  \caption{}
  \label{}
\end{subfigure}%
\begin{subfigure}{.5\textwidth}
  \centering
  \includegraphics[width=\linewidth]{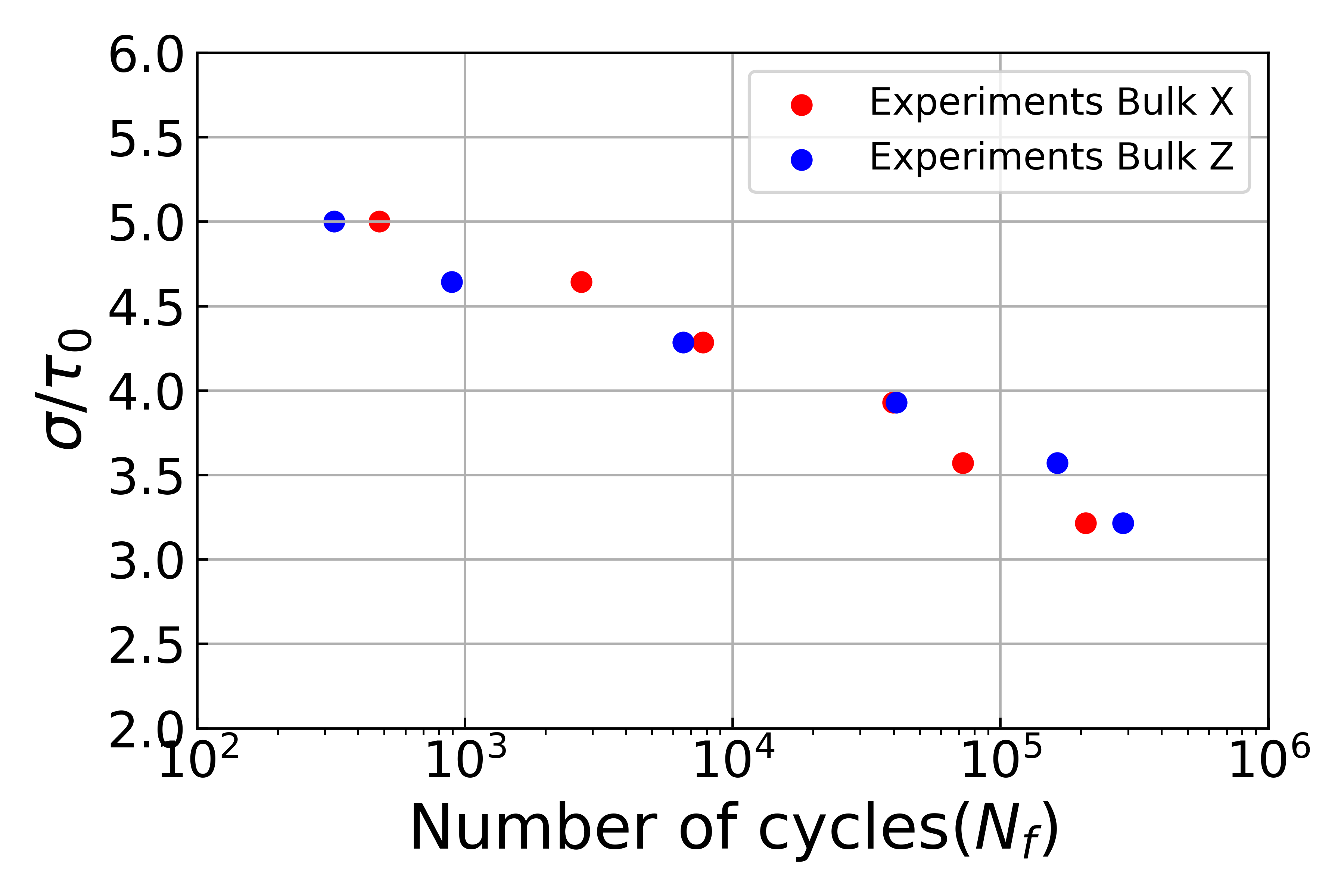}
  \caption{}
  \label{}
\end{subfigure}
\caption{\emph{The experimental fatigue (a)strain-life curve obtained using strain control loading for Bulk X and (b)stress-life curve obtained using stress control loading for Bulk X and Bulk Z samples. }}
\label{fig:SN_curve}
\end{figure}

The fatigue {$\varepsilon$-N} curve obtained for strain control and  {S-N} curve for stress control are represented in Fig.\ref{fig:SN_curve}(a) and Fig. \ref{fig:SN_curve}(b), respectively. In these curves, the fatigue life $N$ corresponds to the cycle in which the final fracture is reached on the specimen.  The fatigue performance of the samples built in the X and Z directions \ref{fig:SN_curve}(b) is different, especially for high stress levels where Bulk X samples present longer fatigue life. This result is in agreement with previous studies by Wang et al.\cite{Wang2011} where the better performance of the samples built in the X direction is also observed for high stresses.

\begin{figure}[h]
	\centering
		\includegraphics[width=1\textwidth]{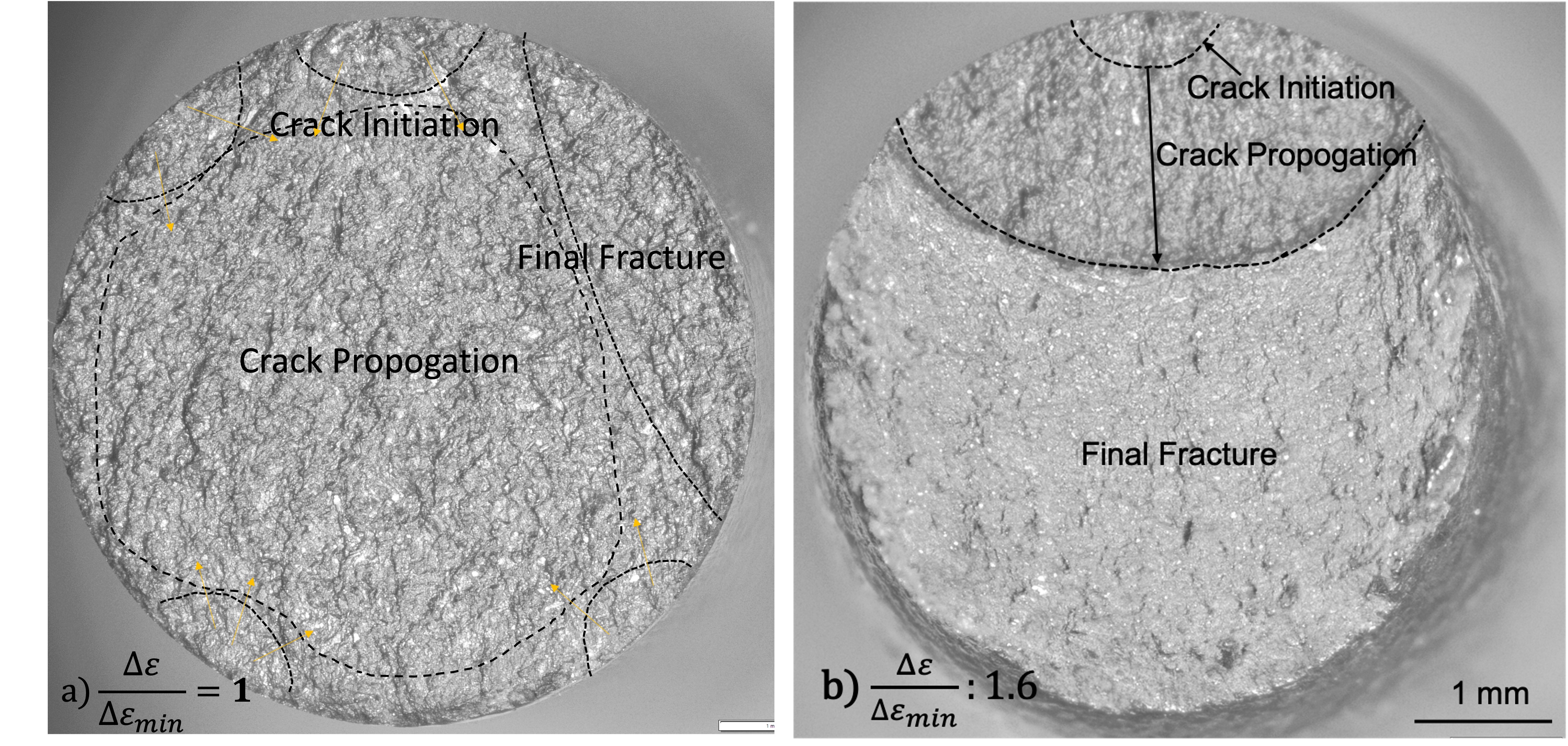}
	\caption{\em{ An overall view of the specimen subjected to cyclic strain amplitude, ${\Delta \epsilon}/{\Delta \epsilon_{min}}$= a) 1 and b) 1.6,  $\%s$ showing crack initiation, propagation, and final fracture regions. 
 }}
	\label{fig:fracture}
\end{figure}

To observe the mechanism of fatigue failure, a fractography analysis was performed for the two samples loaded with cyclic strain amplitude ${\Delta \epsilon}/{\Delta \epsilon_{min}}$, 1 and 1.6 at $750^{\circ}$C.  The fracture surfaces obtained are shown in Fig.\ref{fig:fracture}(a), and Fig.\ref{fig:fracture}(b) respectively, showing the crack initiation (flat areas origin of propagating cracks), propagation (beach marks), and final fracture regions (dimples). It is observed that the dominant crack initiation sites were subsurface areas that have some small defects (both unmelted particles or very small pores have been observed), as they act as stress concentration sites. For low strain amplitude ${\Delta \epsilon}/{\Delta \epsilon_{min}}$=1, multiple crack initiation sites were observed, while at higher applied strains,${\Delta \epsilon}/{\Delta \epsilon_{min}}$=1.6, a single and well-defined crack initiation site was observed. 

Multiple crack initiation sites for small strain amplitude (${\Delta \epsilon}/{\Delta \epsilon_{min}}$=1) implies that very small heterogeneities near the surface are enough to nucleate a crack and therefore the expected scatter in these cases is greater.

The number of cycles for propagation can be qualitatively estimated by the number of beach marks observed in the propagation area. The crack propagation area for both strain ranges is similar, but since the width of the beach marks is smaller for the lower strain amplitude, the propagation regime is slightly larger for the smaller strain range. A rough estimation of the number of propagation cycles is around $0.01N_f$ for ${\Delta \epsilon}/{\Delta \epsilon_{min}}$ = 1 and $0.1N_f$ for ${\Delta \epsilon}/{\Delta \epsilon_{min}}$=1.6, where $N_f$ is the experimental fatigue life. In both cases, the ratio of propagation to total life is small and nucleation can be considered as the most dominant fatigue mechanism for both applied strains. This fact validates the modeling approach followed, in which fatigue life is estimated based on the number of cycles for nucleation.

\begin{figure}[h]
	\centering
		\includegraphics[width=1\textwidth]{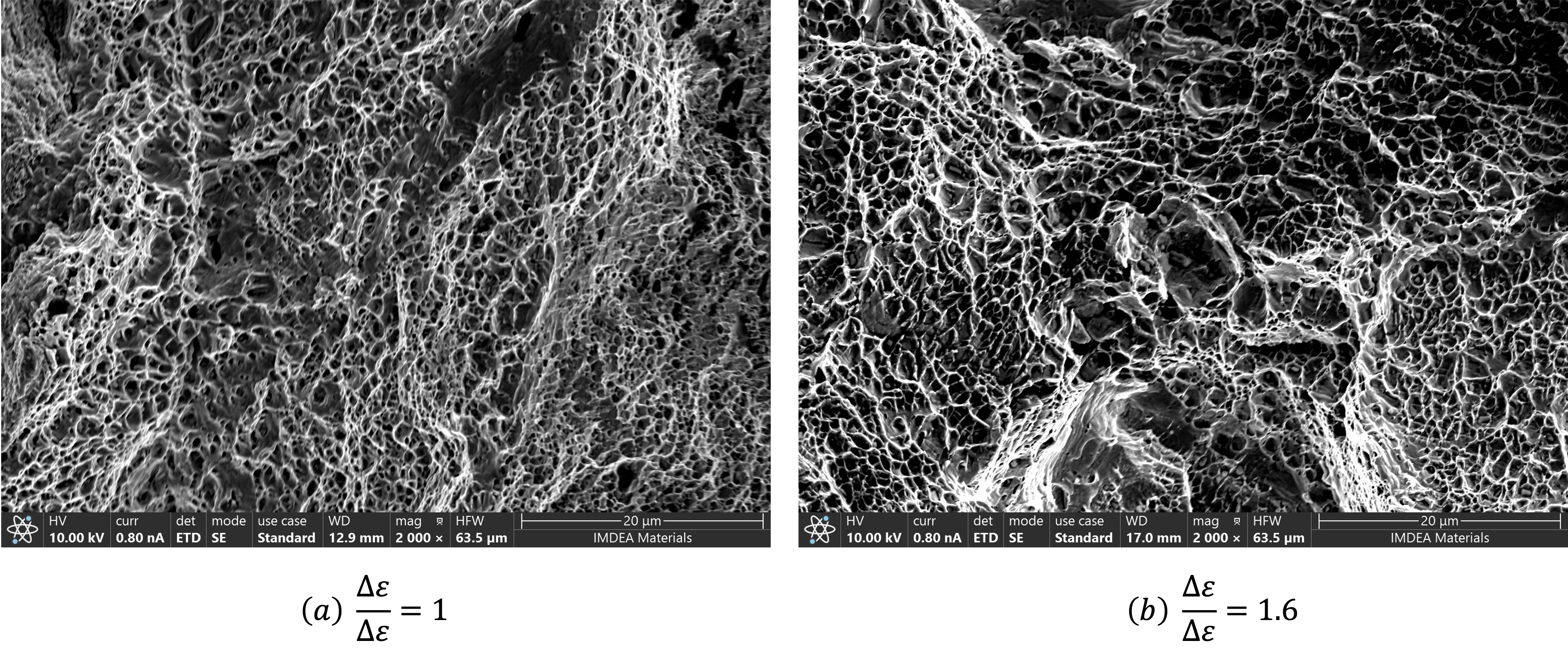}
	\caption{\em{ The final fracture surface of samples subjected to cyclic strain amplitude, ${\Delta \epsilon}/{\Delta \epsilon_{min}}$= a) 1 and b) 1.6,  $\%s$ showing ductile failure with dimples. 
 }}
	\label{fig:duct_fracture}
\end{figure}

The final fracture of the samples was reached when the crack reached the critical size, and the fracture surface area corresponding to this stage was easily observed as a typical ductile fracture surface  with dimples ( Fig.\ref{fig:duct_fracture}). 

\subsection{Numerical results}
The CP-FFT framework presented is applied to predict the cyclic response of bulk SLM Hastelloy-X built in different orientations, thus presenting different microstructural characteristics.

The first objective is to obtain the crystal plasticity parameters for an accurate prediction of the cyclic response for all of the specimens considered. Second, the fatigue life prediction model will be adjusted based on two fatigue tests under strain control. The resulting fatigue life prediction tool will be validated for the rest of the strain-controlled fatigue tests and also for the stress-controlled loading. Finally, the effect of the microstructures resulting from the different SLM processes on fatigue performance will be analyzed.

\subsection*{Crystal plasticity model parameters}
The parameters of the CP model of SLM Hastelloy-X at $750^{\circ}$C after annealing and thermal processing will be obtained. The elastic constants of Hastelloy-X single crystals are taken from Canistraro et al.\cite{canistraro1998elastic}. The strain rate sensitivity coefficient and exponent, $\dot{\varepsilon}_0$ and $m$, respectively, are taken from previous work in Hastelloy-X \cite{Pilgar2022}. The parameters defining the critical resolved shear stress and the evolution of the back stress, which are responsible for isotropic and kinematic hardening, respectively, are obtained by inverse fitting. The input data is the experimental uniaxial cyclic stress-strain response of the bulk sample X loaded in the X direction with strain control and ${\Delta \epsilon}/{\Delta \epsilon_{min}}$= 1.6, the one with the highest strain amplitude. 
The RVE used for the identification of CP parameters is generated to be statistically representative of the actual microstructure (same grain size, aspect ratio, and orientation distributions). The RVE accommodates approximately 1000 grains and is discretized in 64$^3$ voxels. The inverse optimization framework proposed by Herrrera-Solaz et al.\cite{HerreraSolaz2014}, based on the Levenberg-Marqadt least squares algorithm, is used in this study.

Experimentally, it is observed that for SLM-Hastelloy-X, the cyclic stress-strain behavior stabilizes approximately after 2000 cycles for strain-controlled loading at $750^{\circ}$C. Numerically, simulating 2000 cycles makes the optimization process computationally too expensive. Some acceleration techniques have been proposed in the literature, such as the cycle jump approach \cite{Cruzado2017}. However, since the microstructure-sensitive fatigue crack initiation methodology is based on extrapolating the FIP distribution in the stable hysteresis loop, in this work we have opted for a simpler alternative. In our approach, the CP parameters controlling isotropic and kinematic hardening evolution have been fitted such that 15 numerical cycles of the RVE provide a stable hysteresis cycle equal to the experimental stable cycle, reached after 2000 cycles. The number of 15 cycles has been chosen large enough to ensure a progressive stabilization of the cyclic response and local history fields and FIPs. 

\begin{figure}
	\centering
		\includegraphics[width=0.6\textwidth]{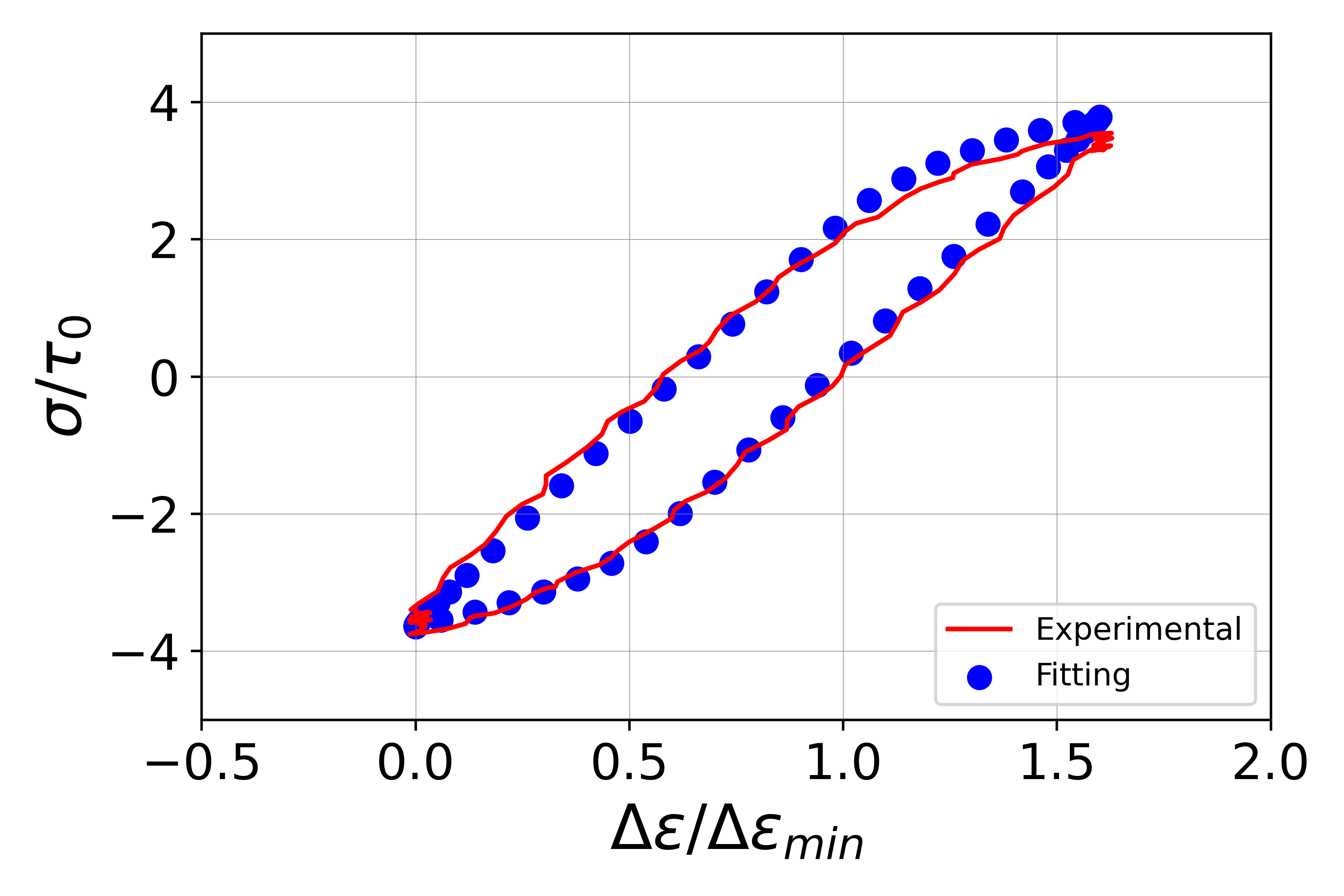}
	\caption{\em{ A stable cyclic experimental loop for applied strain range, ${\Delta \epsilon}/{\Delta \epsilon_{min}}$=1.6 and numerically fitted cyclic loop using LM algorithm.  }}
	\label{fig:fit_Inverse}
\end{figure}

The CP parameters correspond to the Voce isotropic hardening   
 ($\tau_0, \tau_s$, $h_0$ and $h_s$), and the kinematic hardening of the Ohno-Wang model ($c, d, k$) are identified using the procedure mentioned above, and these values are tabulated in Table \ref{tab:CP_paramters}
 along with the remaining CP parameters. The stable experimental cyclic stress-strain curve of the 2000th cycle is represented in Figure \ref{fig:fit_Inverse} together with the polycrystalline cyclic response of the 15th cycle after adjustment of the CP parameters. The figure shows an almost superimposed response between the model and the experiment, confirming the accuracy of fitting the parameters of the CP model.
\begin{table}[htbp]
	\centering
	\caption{Crystal plasticity parameters  of a SLM Hastelloy-X at {$\mathbf{750^{\circ}}$\textbf{C}} }
    \vspace{1mm} 
	
	\begin{tabular}{lllll}
	\hline
		Elastic $(GPa)$ & $\text{C}_{11}$ & $\text{C}_{12}$& $\text{C}_{44}$ & \\
	
		&$\text{194.8}$ & $\text{126.91}$ & $\text{89.95}$&\\
	
		Viscoplastic & $m$ &${\dot{\gamma}}_{0}(s^{-1})$& &  \\
		& $\text{0.017}$ & $\text{2.42}\times {\text{10}}^{-3}$ & & \\
\bf{Isotropic Hardening (MPa)} & $\boldsymbol{\tau_{0}} $& $\boldsymbol{\tau_s}$ & $\boldsymbol{h_0}$ & $ \boldsymbol{h_s}$\\
		& $\boldsymbol{\tau_{0}}$ & $\boldsymbol{\textbf{1.25}\tau_{0}}$ & $\textbf{78.7} \boldsymbol{\tau_{0}}$ & $\textbf{0.71} \boldsymbol{\tau_{0}}$\\
		\bf{Kinematic Hardening} & \bf{$\textbf{c}$(MPa)} & $\boldsymbol{d}$ & $\boldsymbol{k}$\\
		& $\mathbf{171.42 \tau_{0}}$ & $\mathbf{2.57\tau_0}$ & $\mathbf{0.1066}$\\

	    \hline
	\end{tabular}
	\label{tab:CP_paramters}
\end{table}

\subsection{Prediction of cyclic plastic response}
The ability of the set of CP parameters fitted above to predict the cyclic deformation behavior for different applied strain amplitudes is presented in this section. 
In particular, the stress-strain curves and the evolution of minimum and maximum stress with the number of real cycles are predicted under the uniaxial cyclic strain amplitude $\frac{\Delta \epsilon}{\Delta \epsilon_{min}}$= 1, 1.1, 1.2, 1.4, 1.6, 1.8,  with $R_\epsilon=0$. All the cases considered correspond to bulk-X samples. 

\begin{figure}
\begin{subfigure}{.5\textwidth}
  \centering
  \includegraphics[width=\linewidth]{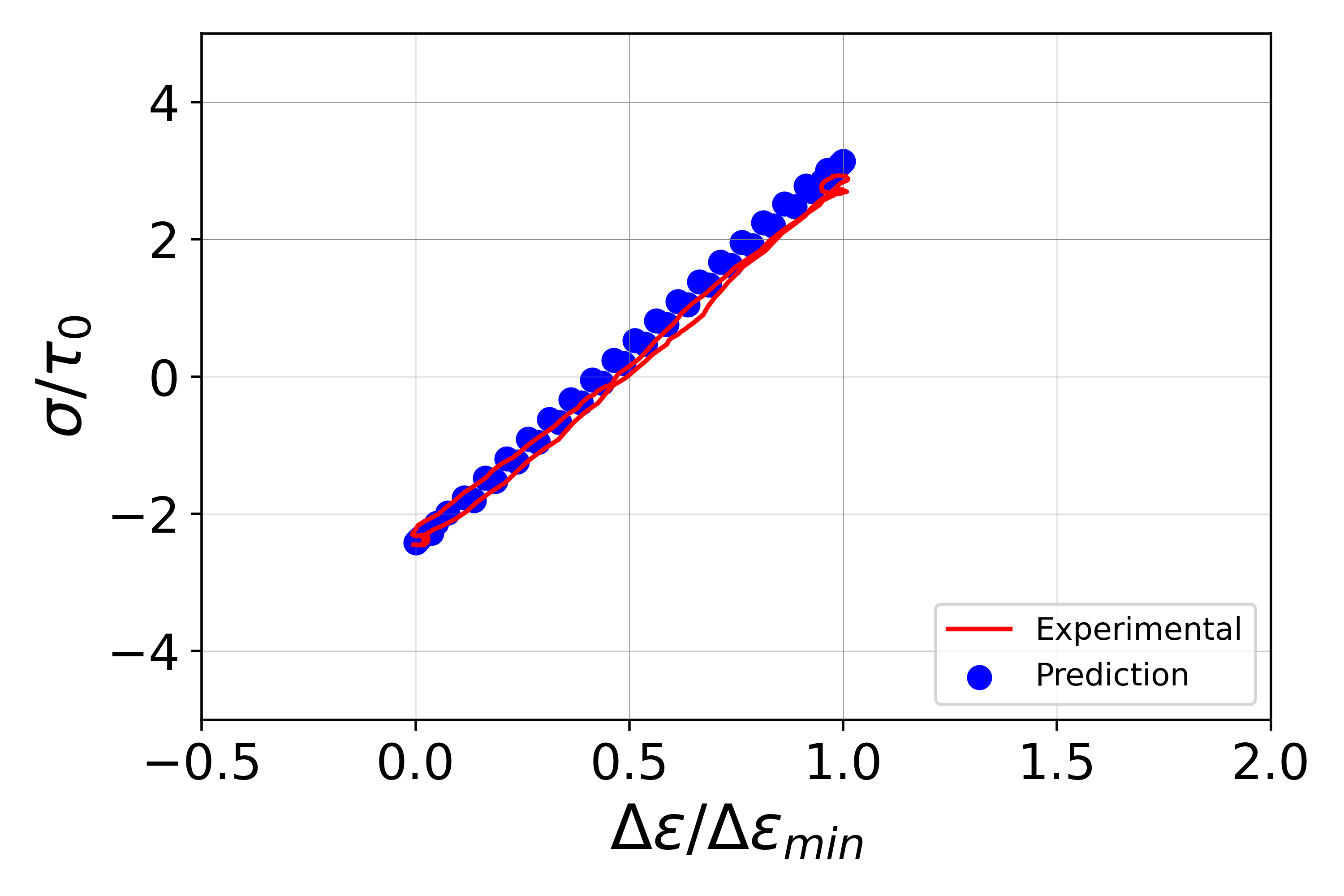}
  \caption{}
\end{subfigure}%
\begin{subfigure}{.5\textwidth}
  \centering
  \includegraphics[width=\linewidth]{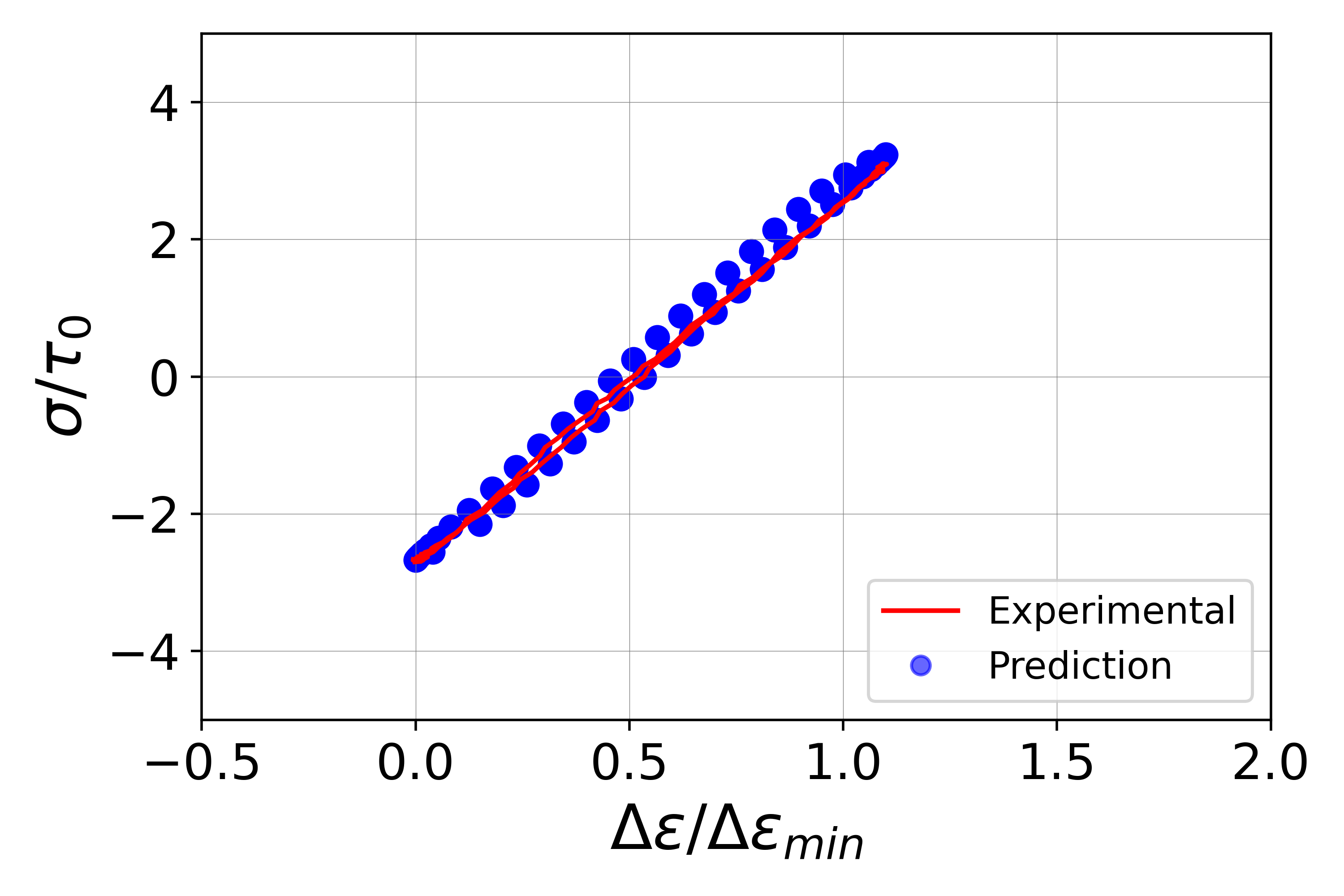}
  \caption{}
\end{subfigure}
\begin{subfigure}{.5\textwidth}
  \centering
  \includegraphics[width=\linewidth]{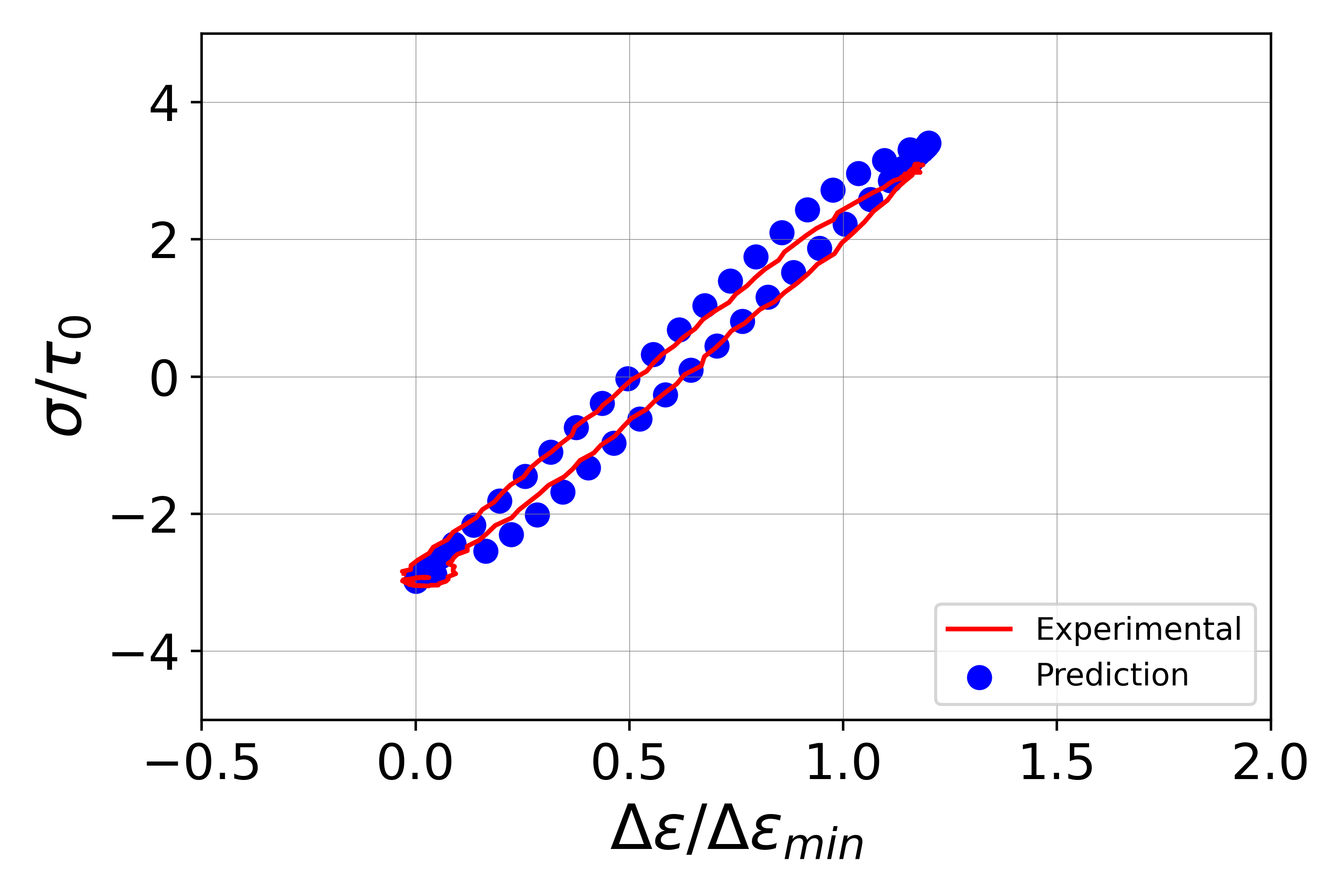}
  \caption{}
\end{subfigure}
\begin{subfigure}{.5\textwidth}
  \centering
  \includegraphics[width=\linewidth]{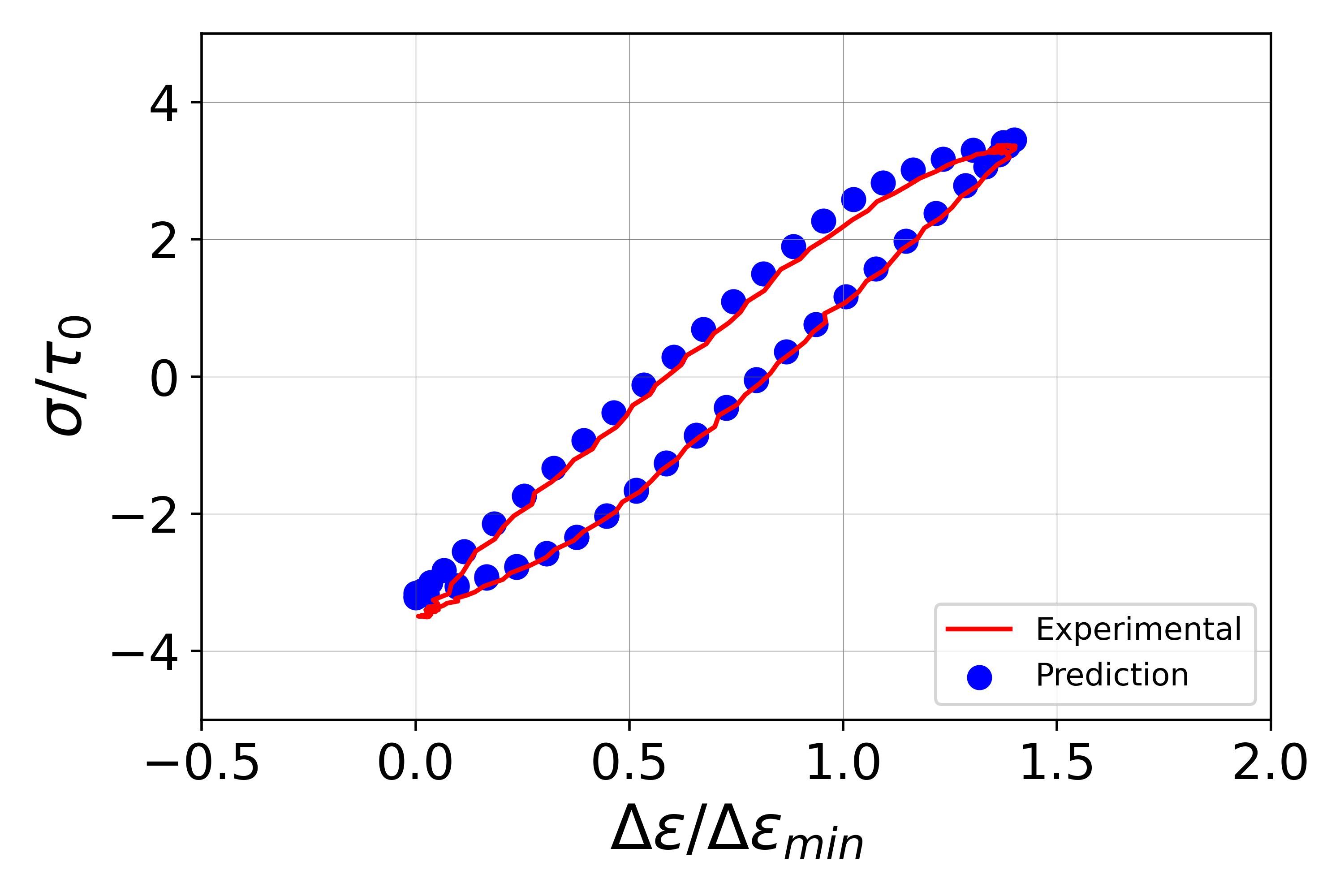}
  \caption{}
\end{subfigure}
\begin{subfigure}{.5\textwidth}
  \centering
  \includegraphics[width=\linewidth]{0_4_pred}
  \caption{}
\end{subfigure}
\begin{subfigure}{.5\textwidth}
  \centering
  \includegraphics[width=\linewidth]{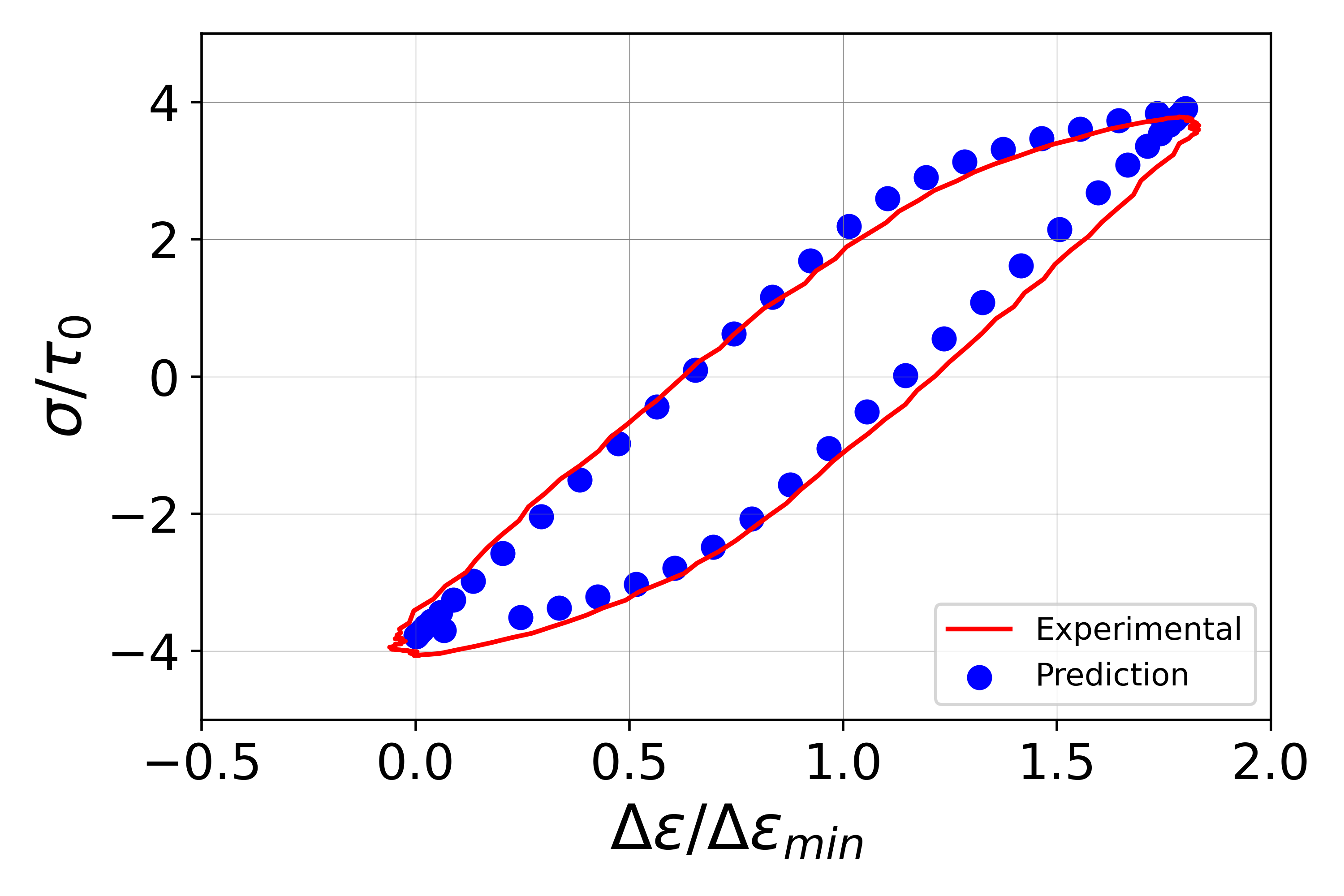}
  \caption{}
\end{subfigure}
\caption{\emph{Comparison between stable experimental 2000th cycle (solid red lines), and numerical cyclic 15th cycle (dotted blue lines)  hysteresis loops for different applied strain ranges,(${\Delta \epsilon}/{\Delta \epsilon_{min}}$= a)1, b)1.1, c)1.2, d)1.4, e)1.6, and f)1.8. Note that $\tau_0$ normalizes stresses. }}
\label{fig:All_hyst}
\end{figure}

The stable experimental hysteresis curve, corresponding to the 2000th experimental cycle, is represented in Fig. \ref{fig:All_hyst} for six different strain amplitudes together with the computational homogenization results for the 15th cycle. The comparison shows that an accurate stabilized cyclic stress-strain loop can be predicted for all the applied strain amplitudes. These results validate the simulation strategy, based on accelerating the hardening evolution, to predict the stable cycle under a wide range of cyclic loading conditions.

\begin{figure}
\begin{subfigure}{.5\textwidth}
  \centering
  \includegraphics[width=1\linewidth]{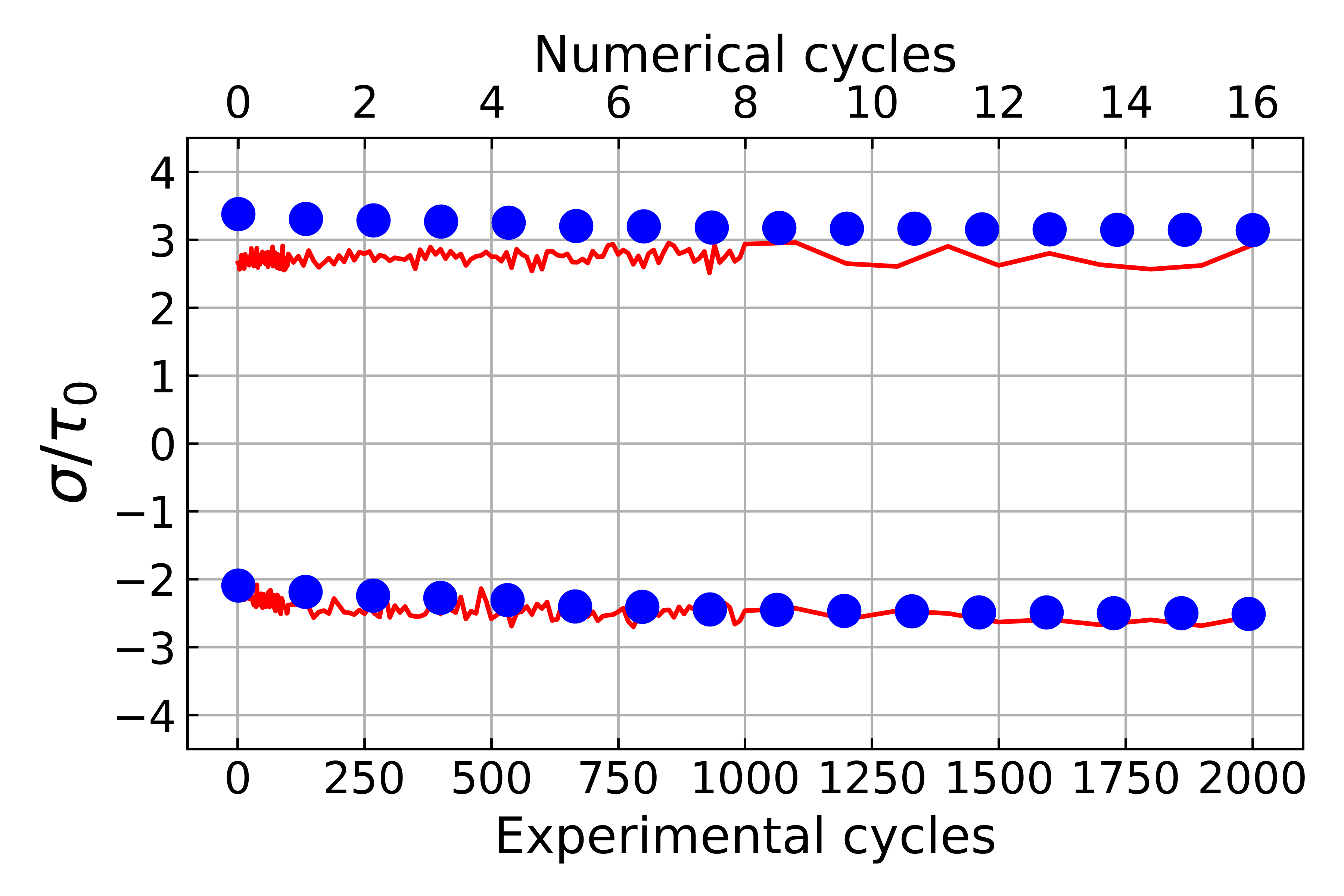}
  \caption{}
\end{subfigure}%
\begin{subfigure}{.5\textwidth}
  \centering
  \includegraphics[width=\linewidth]{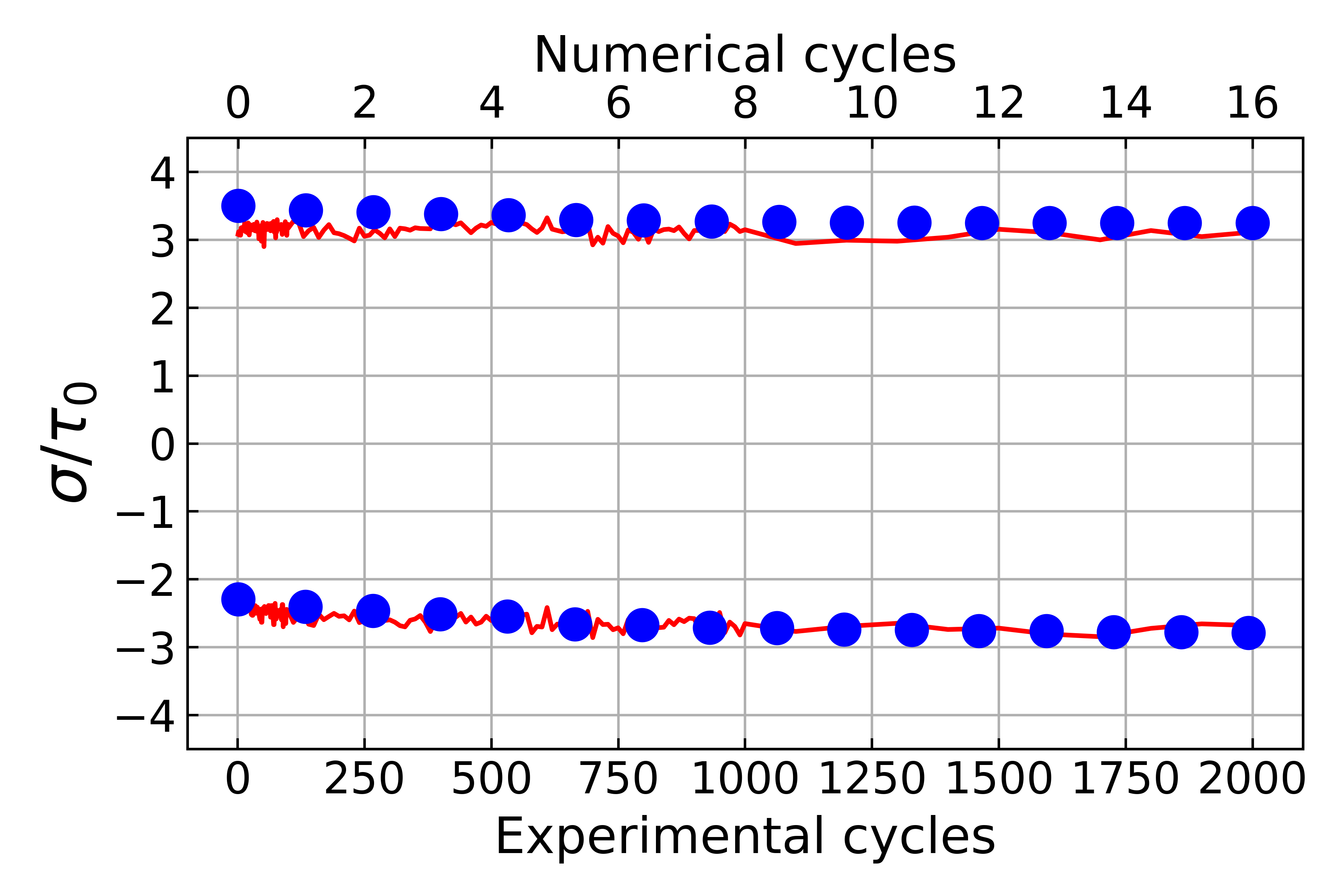}
  \caption{}
\end{subfigure}
\begin{subfigure}{.5\textwidth}
  \centering
  \includegraphics[width=\linewidth]{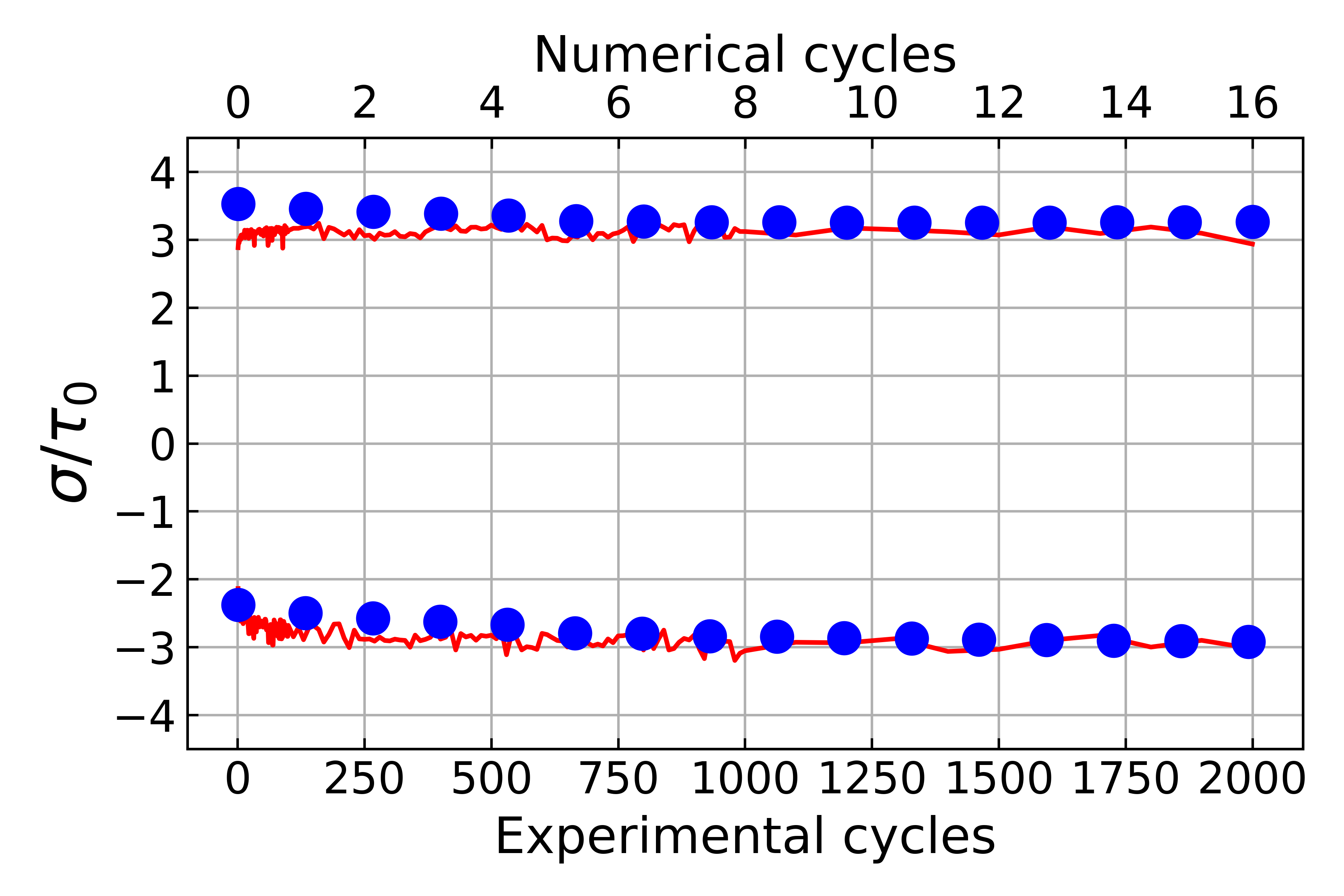}
  \caption{}
\end{subfigure}
\begin{subfigure}{.5\textwidth}
  \centering
  \includegraphics[width=\linewidth]{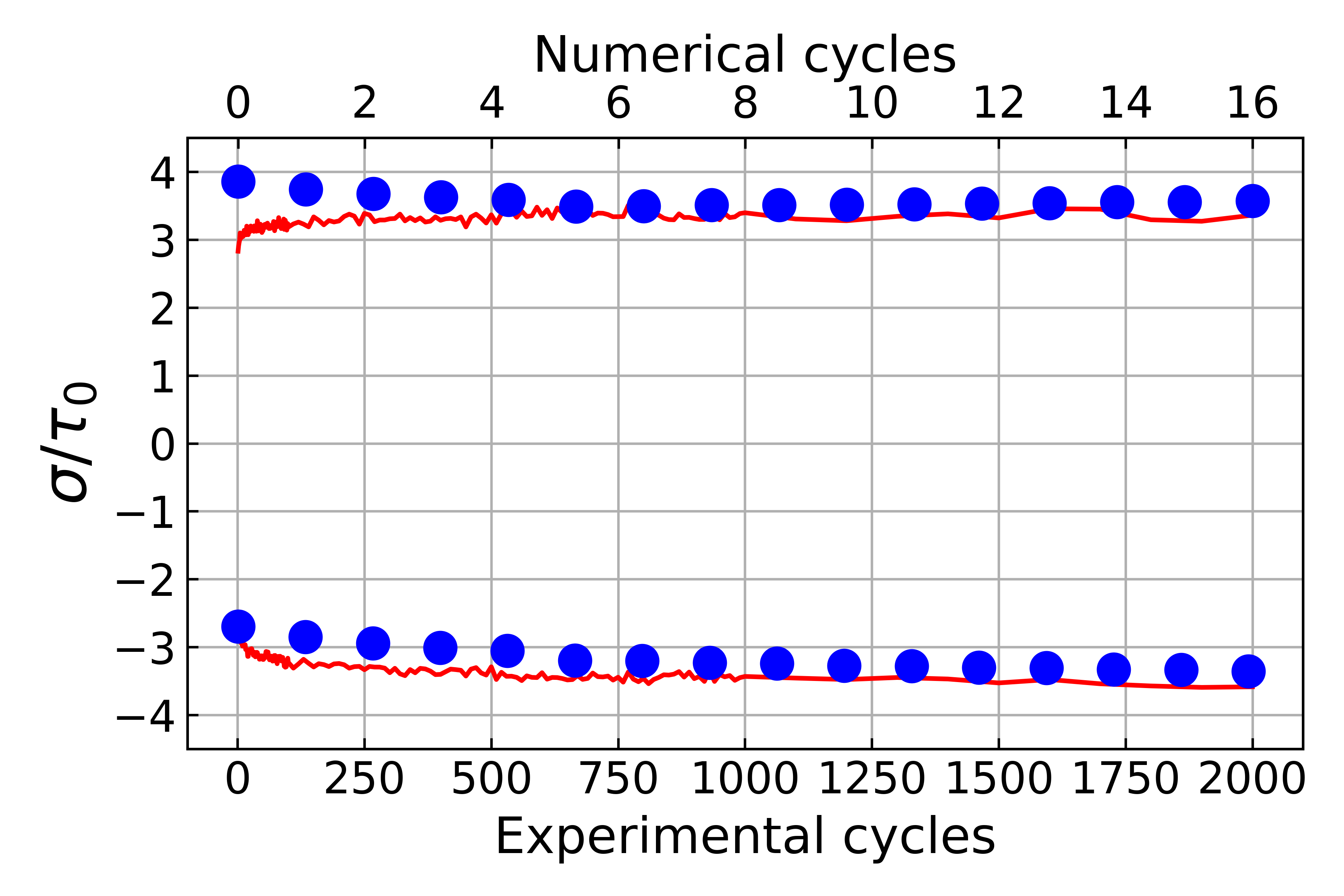}
  \caption{}
\end{subfigure}
\begin{subfigure}{.5\textwidth}
  \centering
  \includegraphics[width=\linewidth]{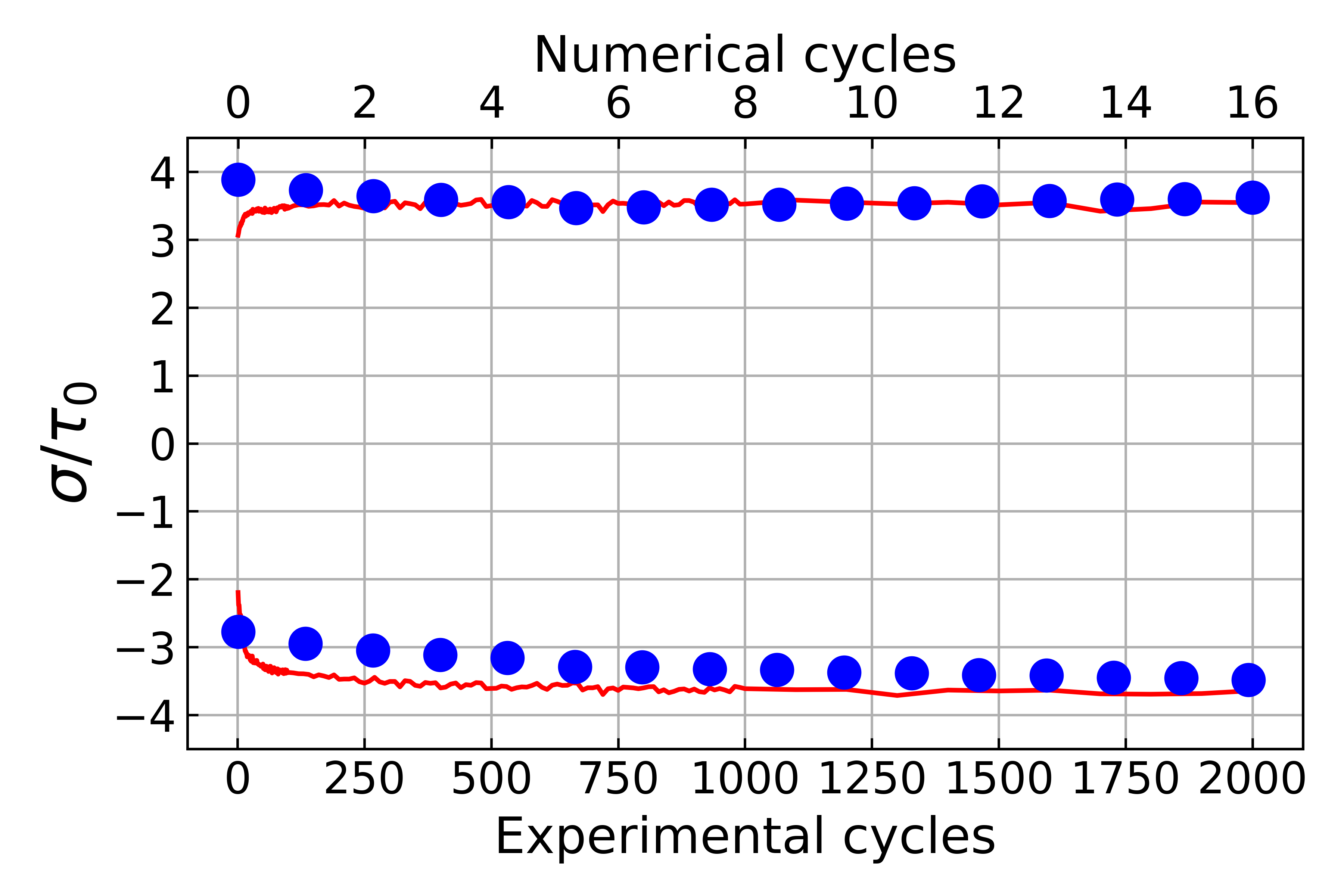}
  \caption{}
\end{subfigure}
\begin{subfigure}{.5\textwidth}
  \centering
  \includegraphics[width=\linewidth]{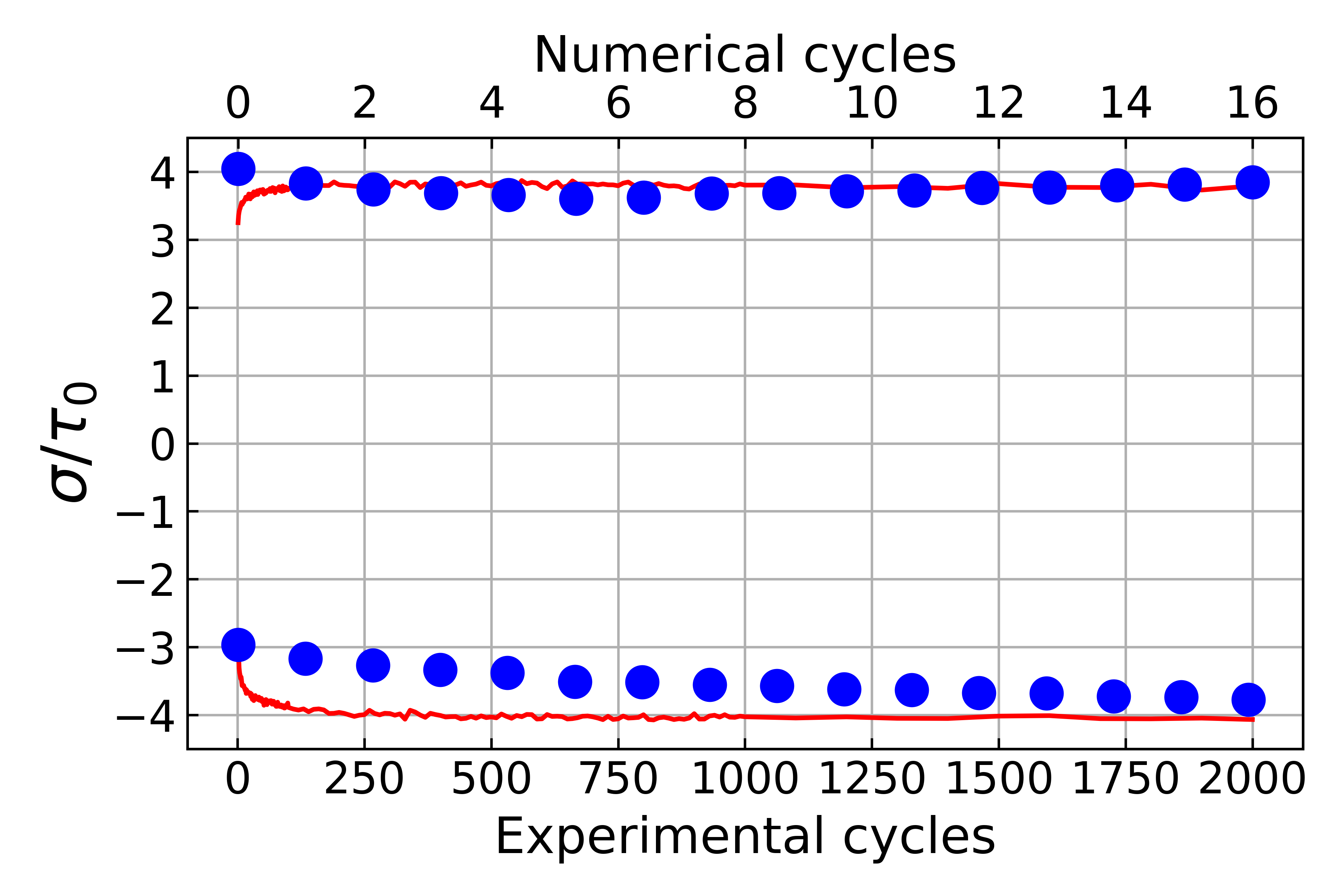}
  \caption{}
\end{subfigure}
\caption{\emph{Comparison between experimental(solid red lines) and numerical cyclic(dotted blue lines) response based on the evolution of maximum and minimum stresses with the number of cycles for different applied strain ranges,(${\Delta \epsilon}/{\Delta \epsilon_{min}}$= a)1, b)1.1, c)1.2, d)1.4, e)1.6, and f)1.8. Note that $\tau_0$ normalizes stresses. }}
\label{fig:Peak_stresses}
\end{figure}

The evolution of peak stresses until cycle stabilization is represented in Fig.\ref{fig:Peak_stresses} for both the experimental results and the numerical predictions. { The experiments show an initial asymmetry in tension/compression maximum stresses that progressively disappears  and that is also captured by the simulations. It can be observed that the experimental results always stabilized before 2000 cycles and the numerical value for 15 cycles coincides with the stable value for all the strain range. The evolution of the stress peaks obtained by the model before stabilization also follows the experimental trend. The comparison of experimental and predicted peak stresses in Fig.\ref{fig:Peak_stresses} show a considerably larger difference (up to $20\%$) in the initial cycles with respect the rest of the cycles. The reason is that for large applied strains a fast decay in the maximum stresses is observed in the experiments for a small number of cycles, and the proposed acceleration, where each model cycle corresponds to 125 cycles, cannot reproduce that decay preserving the accuracy for the more gradual cases. Nevertheless, the difference between peak stresses at the 15th numerical cycle and the 2000th experimental cycle for tensile loading is always less than $6\%$ and less than $8\%$ for compressive loading, even for the large applied strain cases where the initial decay was spread in more cycles. This small differences in all the very different cases considered show a remarkable accuracy of the model in predicting stabilized cyclic response with the accelerated technique.}

\subsection{Fatigue Life Prediction}
\subsubsection{Life prediction model calibration}
The parameters of the fatigue life models, $W_{crit}$ and $m$ in Eq.\ref{eq16} and $W^{crit}_\sigma$ and $n$ in Eq.\ref{eq17}, respectively, for the strain- and stress-controlled loading, must be calibrated using some of the experimental data. In each case, two different experimental results and their corresponding cyclic behavior are necessary to predict the power law constants in Eqs.\ref{eq16} and \ref{eq17}. Furthermore, to consider the stochastic nature of fatigue, the average response of the simulation of 20 different RVE realizations (also known as Statistical Volume Element, SVE) was considered for each applied load. The use of an SVE is necessary to consider the differences between the results of different RVEs statistically. { The fatigue model parameters obtained for the strain-controlled loading and stress-controlled loading are tabulated in Table \ref{tab:fatigue_para}.}

\begin{table}[htbp]
	\centering
	\caption{The fitted parameters for the proposed fatigue crack initiation models  }
	\vspace{1mm} 
	\label{tab:fatigue_para}
	\begin{tabular}   {lllll}
	\hline
	   &  \multicolumn{2}{c}{$\varepsilon$ - controlled loading}    &  \multicolumn{2}{c}{$\sigma$ - controlled loading}\\
	 \hline
		{Parameter}  & $W_{crit}$ & $m$ & ${W_{\sigma}}^{crit}$ & $nk$  \\
		{Value}  & $3.421 \times 10^{19}$ & $2.62$ & $ 4.06 \times 10^8$  & $-0.56$  \\
		\hline
	\end{tabular}
\end{table}

\subsection{Fatigue Life Prediction: strain controlled tests}

\begin{figure}[htbp]
	\centering
		\includegraphics[width=0.8\textwidth]{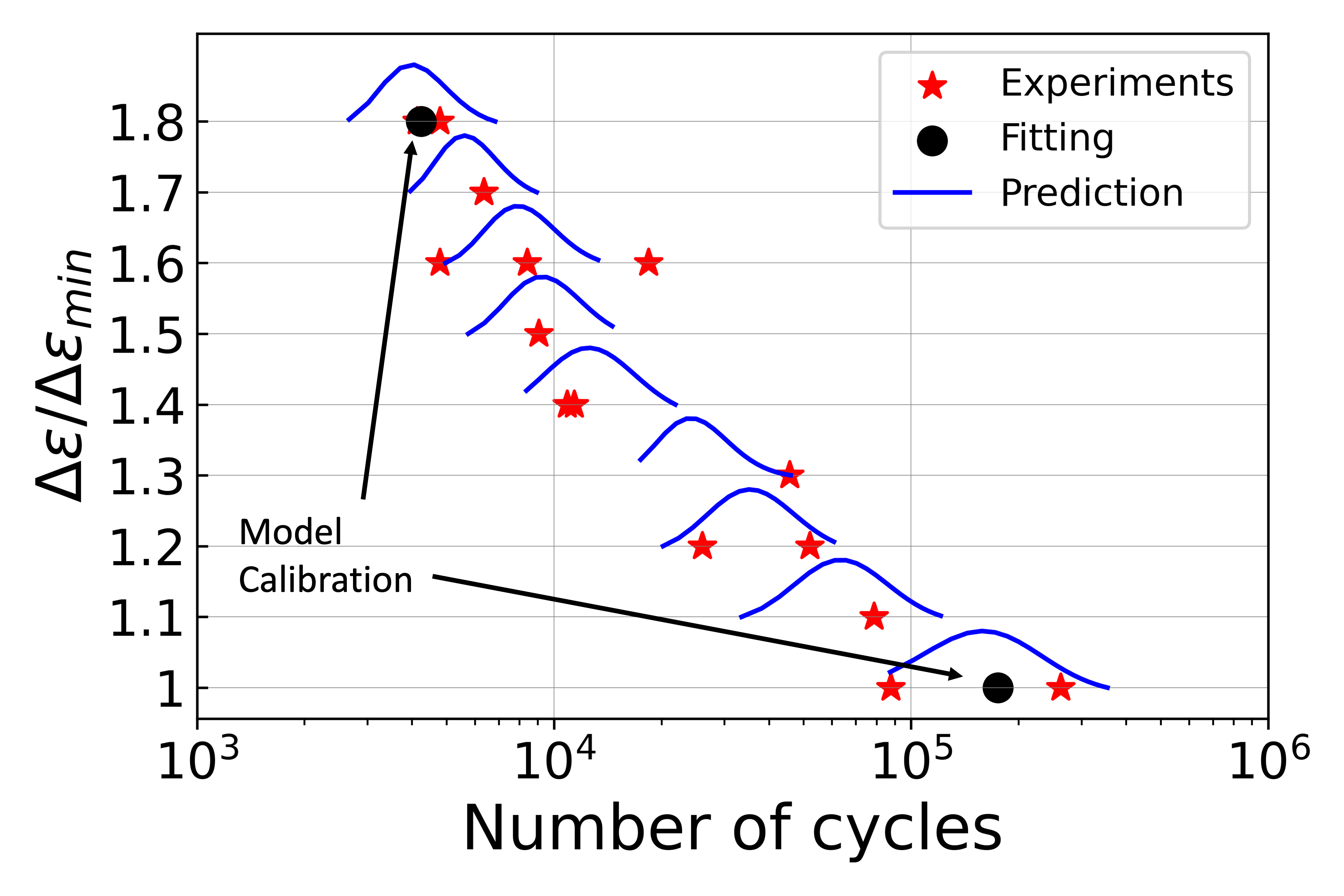}
	\caption{\em{ Experimental fatigue life and model predictions for bulk X under strain-controlled loading at  $750^{\circ}$C. Note that the minimum cyclic strain amplitude $\Delta\epsilon_{min}$ normalizes strain. }}.
	\label{fig:strain_life}
\end{figure}

{ The cyclic stable FIPs and fatigue model in Eq. \ref{eq16} are used to predict fatigue life as a function of applied strain and microstructure. The experimental fatigue lives obtained for different experiments at each strain amplitude are plotted in Fig. \ref{fig:strain_life} using red stars.  The average experimental results for two strain ranges $\frac{\Delta \epsilon}{\Delta \epsilon_{min}}$= 1 and 1.8, marked with black dots,  and the average FIP of the simulation of 20 RVEs for these two applied strain ranges are used to calibrate the fatigue model parameters. Figure \ref{fig:strain_life} represents, using small blue curves near each experimental point, the fatigue life probability distribution predicted by using the results of the 20 RVEs, for each strain amplitude.}
Very good agreement between the experiments and simulations is observed for all applied cyclic strain amplitudes. { When comparing the average experimental fatigue life with the peak of probability of failure provided by the model distributions, the errors are below 20\% except for the case of  $\frac{\Delta \epsilon}{\Delta \epsilon_{min}}$= 1.3, where the error grows to 61\%. Moreover, the average experimental life lied in all the cases in the percentile 95\% of the model predictions, with the exception of the 1.3 case. The experimental life for this case breaks the tendency observed in the experimental results, and is longer than for a lower strain range of 1.2. Model cannot capture this abnormal tendency that is probably caused by some experimental issue.}

Furthermore, the results highlighted the predictive capacity of the model to consider the change in the FIP distribution between high and low strain amplitudes. The distribution of plastic deformation is heterogeneous at lower strain amplitudes, and very few grains undergo plastic deformation. However, at higher strain amplitudes, more homogeneous plastic deformation was observed. The effect of microstructure has a more pronounced effect at lower strain amplitudes than at higher strain amplitudes, since the localized continuous accumulation of plastic slip in very few grains leads to considerable experimental and numerical scatter in the fatigue life. 

\subsection{Fatigue Life Prediction: Stress Controlled tests}
The crystal plasticity model presented in section \ref{crystal_plasticity} is extended to predict fatigue life under stress-controlled loading. In this case, tests are available in two different building directions, Bulk X and Bulk Z, so the predictive capacity of the model for different microstructures will be assessed.

The methodology to predict fatigue life under stress-controlled loading is described in Section \ref{stress_controlled}. Cumulative cyclic damage in the plastic regime ($W_p^{acc}$) is specified as the precursor to the onset of fatigue cracks, and fatigue live $N$ is predicted using the modified criteria (Eq.\ref{eq17}-\ref{eq19}). Since under stress control, the scatter is usually larger, 50 RVEs were generated with an average grain size, aspect ratio, and grain orientations for Bulk X and Bulk Z samples. Virtual uniaxial cyclic tests were performed on RVEs under stress-controlled loading with ${\Delta \sigma}/{\tau_0}$= 3.2, 3.57, 3.93, 4.29, 4.64, and 5 and under a stress ratio $R_\sigma$ = 0.03. Note that $\tau_0$ corresponds to the critical resolved shear stress(CRSS). The adjustment parameters $W^{crit}_{\sigma}$ and $nk$ in Eq.\ref{eq19}, which are necessary to predict the fatigue crack initiation under stress-controlled loading, were identified from the fatigue life obtained in two different tests (${\Delta \sigma}/{\tau_0}= 3.2, 5$) for a unique type of material, the one corresponding to fabrication in the X direction (Bulk X). The values of the fitting parameters are $W^{crit}_\sigma = 4.06 \times 10^8$ and $nk=-0.56$. The accumulated FIP, $W_p^{acc}$ of these tests was obtained from the simulation of a SVE with 50 different RVEs. The large size of the SVE helps predict the stochastic nature of the fatigue life for a given microstructure and loading conditions. 

Once the life prediction model is fitted, the SVE for Bulk-X samples is simulated and used to obtain the fatigue life prediction for the rest of the cases. In the case of Bulk-Z, a new SVE is generated containing 50 statistically representative RVEs of the microstructure of the material, which is different from the one of the Bulk-X samples. This SVE is then used to predict the fatigue life of these samples. It is important to note that no experimental results on Bulk-Z material have been used to calibrate the model, so the results are pure predictions and will assess the validity of the model to consider different microstructures for the same parameters.

\begin{figure}
\begin{subfigure}{.5\textwidth}
  \centering
  \includegraphics[width=1\linewidth]{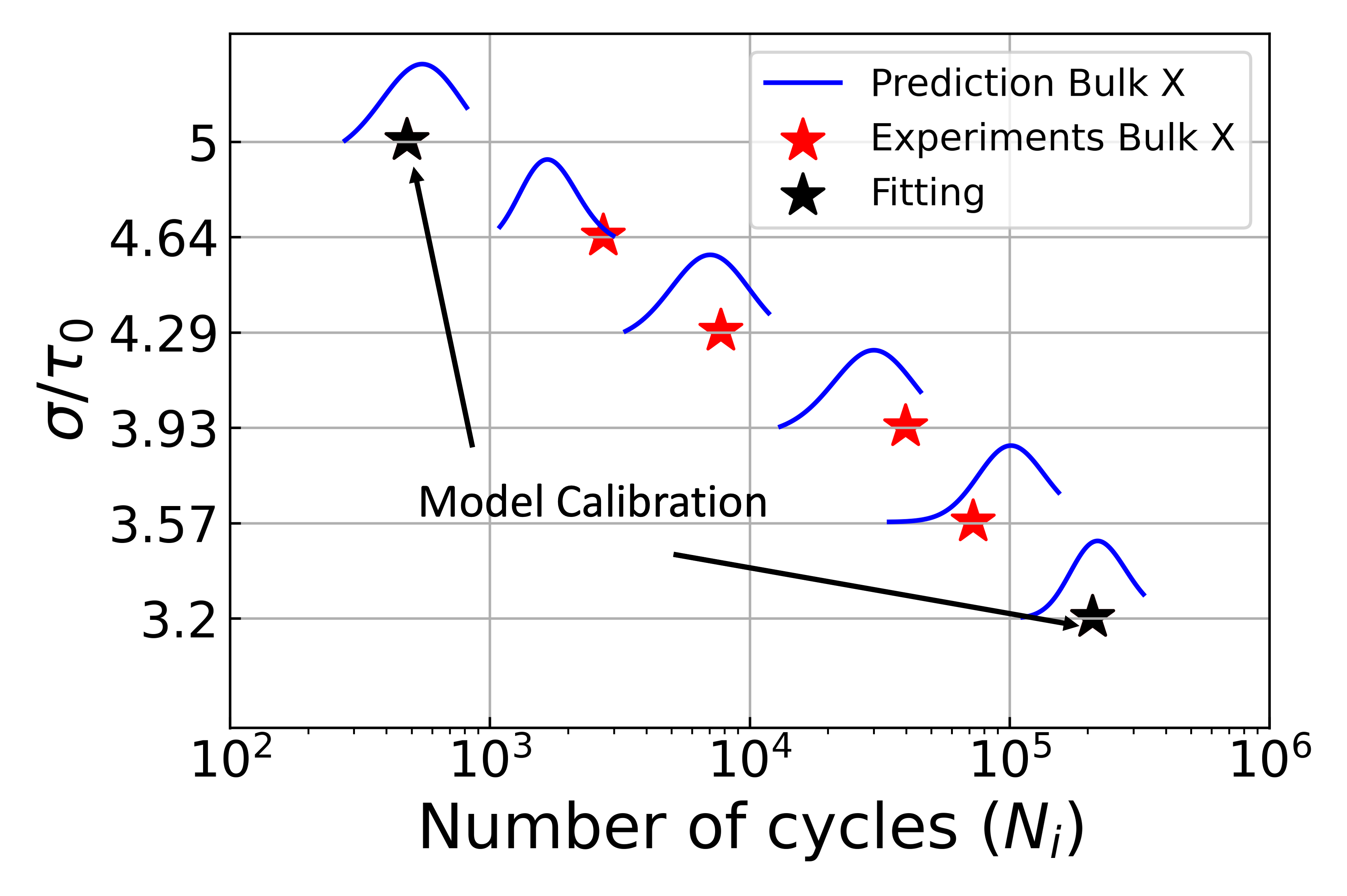}
  \caption{}
  \label{fig:peak_0_25}
\end{subfigure}%
\begin{subfigure}{.5\textwidth}
  \centering
  \includegraphics[width=\linewidth]{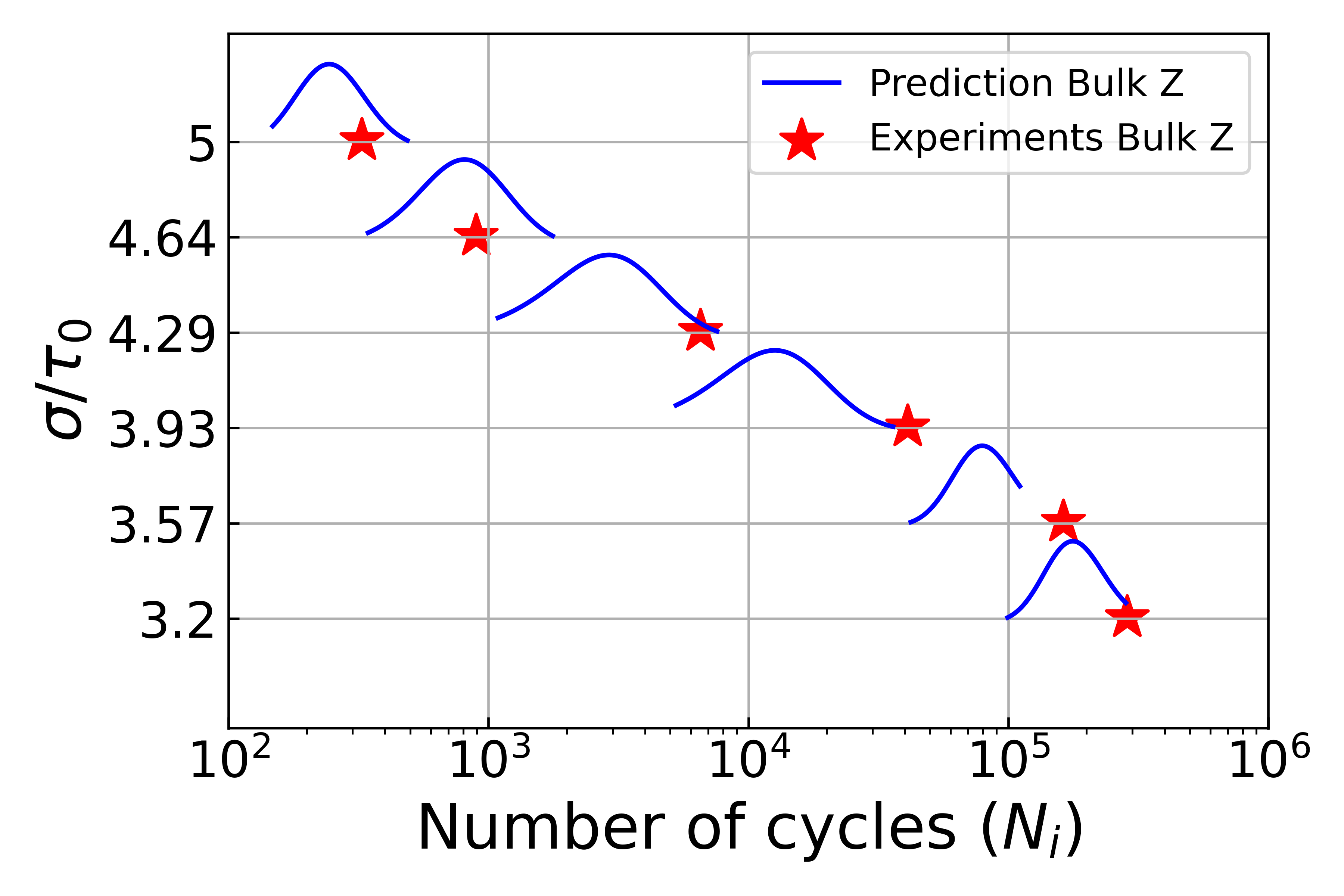}
  \caption{}
  \label{fig:peak_0_275}
\end{subfigure}
\caption{\emph{Experimental results and numerical model predictions fatigue crack initiation life for SLM Hastelloy-X at $750^{\circ}$C as a function of cyclic applied stress,${\sigma}/{ \tau_{0}}$. Note that the stress is normalized by $\tau_0$. a) Bulk X b) Bulk Z. These results are for 50 RVE realizations for each applied stress, with a confidence level of $95\%$. }}
\label{fig:Stress_fatigue}
\end{figure}

The fatigue crack initiation predictions of the model for different applied cyclic stress ranges, ${\Delta \sigma}/{\tau_0}$ are represented in Fig. \ref{fig:Stress_fatigue} for the orientation of the sample, (a) Bulk X and (b) Bulk Z, together with the respective experimental life. 

The agreement between experimental and numerically predicted fatigue life is { very accurate for both building directions in most of the strain ranges studied}. The largest difference was found for the bulk Z sample in the stress ranges ${\Delta \sigma}/{\tau_0}$, 3.57 and 3.93, where fatigue life is slightly underestimated. These results highlight that the model can accurately predict the fatigue life that accounts for the effect of the microstructure. The agreement with the experimental results for the Bulk-Z direction is remarkable because the fatigue life law used for the predictions was calibrated only using two experiments of Bulk-X samples. 

\begin{figure}
	\centering
		\includegraphics[width=0.8\textwidth]{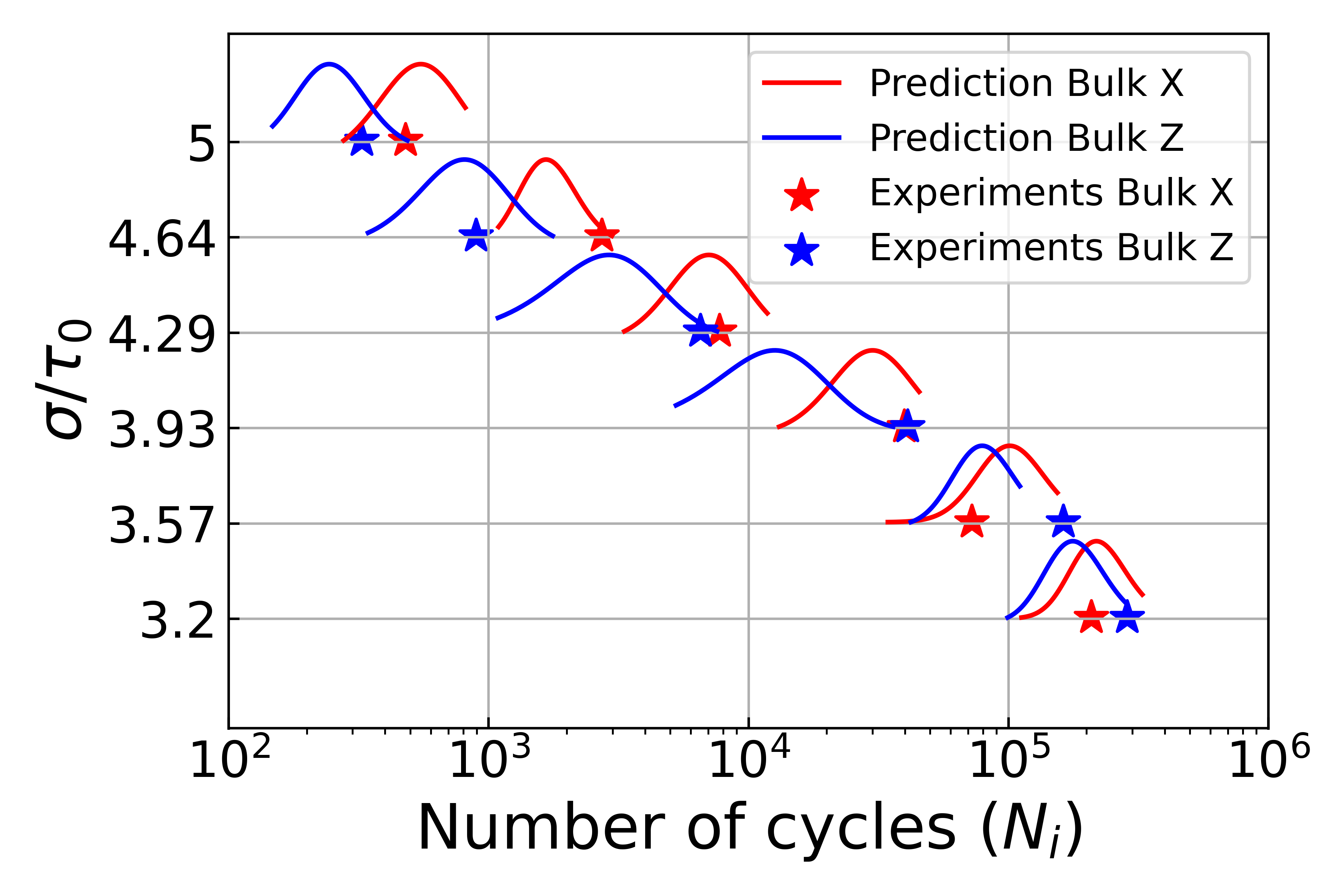}
	\caption{\em{ Comparison between predicted and experimental fatigue lives of Bulk X and Bulk Z, showing the effect of build orientation on fatigue lives. 
 }}
	\label{fig:Compare_life}
\end{figure}

To analyze the effect of the building direction, the S-N curves for the two building directions together are represented in Fig. \ref{fig:Compare_life}. In this figure, it can be observed that the direction of SLM fabrication significantly affects the fatigue life of Hastelloy-X. Experimentally, the horizontally built specimen (Bulk X) shows superior resistance to the initiation of fatigue cracks compared to the vertically built specimen (Bulk Z), with the exception of the two smallest stress ranges applied. These results are in agreement with the observations of Wang et. al. \cite{Wang2011} for the same alloy, Hastelloy-X, and of Bayati et al.\cite{Bayati2020} for SLM Ni-Ti. Furthermore, the better fatigue performance of the printed Bulk-X Hastelloy-X samples corresponds to a stronger quasistatic response, as shown in \cite{Pilgar2022}.

The model can reproduce the better performance of Bulk X compared to Bulk Z for stresses greater than {3.93$\tau_0$} observed in the experiments. Multiple factors could potentially contribute to this anisotropic fatigue performance, such as porosity, residual stresses, surface roughness, and orientation of the deposited layer with respect to the loading direction. However, the fabrication was designed to minimize most of them. For example, minimal porosity was observed in the samples as a result of an optimal selection of the SLM parameters, and a fine machining ensures that the sample has very minimal surface roughness. With respect to the level of residual stresses, it should be very small due to the annealing process and heat treatment. Therefore, the differences observed in the experiments should be mainly caused by the different microstructures (grain size, aspect ratio, and orientation) generated during SLM fabrication in different directions with respect to the loading axis. In the absence of porosity and defects, the slip band precursors of the nucleating cracks tend to appear in grains, whose shape, orientation with respect to macroscopic load, and surrounding area are specially suited for localizing plastic deformation and, therefore, the polycrystalline microstructure plays an essential role. Since the two materials have a very similar texture and average grain size, the main difference in their microstructures is their
grain shape, having Bulk X grains a lower aspect ratio. The more equiaxed grain shape should then provide better resistance to fatigue, because the slip bands are larger and localize more plastic strain in the elongated grains.

Both the model and the experiments agree that the effect of anisotropy in fatigue life increases with the applied stress, being maximal for the highest stress amplitudes, where most of the grains show plastic deformation. At lower stresses, for long LCF lifes, the model shows a similar fatigue endurance for both samples. The prediction of a reduction in fatigue performance anisotropy has also been observed for SLM samples \cite{Wang2011,Bayati2020}. 

 In summary, in addition to the effect of defects and porosity, the orientation of the building influences the lifetime of fatigue.  The S-N curve shows a higher slope for horizontally oriented samples and lower slopes for vertically oriented samples. The reason behind the effect of the building direction on fatigue performance is the difference in the microstructures developed during the process. This influence is strong at higher applied stresses; however, such effects are minimal at lower applied stresses. Therefore, to design the SLM of components against fatigue loading, microstructural aspects (grain aspect ratio, average size, and orientation result of the building orientation should be considered, as they significantly influence fatigue behavior. Furthermore, to capture this fatigue anisotropy, the use of microstructure-sensitive fatigue life nucleation approaches based on polycrystalline homogenization is essential.

\section{Conclusion}
A microstructure sensitive fatigue life prediction framework, based on CP-FFT, is proposed to model the fatigue performance of SLM fabricated Hastelloy-X at $750^{\circ}$C considering the effect of the microstructures developed during the SLM process. The strategy is based on obtaining probability distributions of fatigue indicator parameters (FIP) from the simulation of the cyclic response of a statistical volume element (SVE) formed by several representative volume elements (RVE). Then, the fatigue life is obtained for each particular case from the FIP distribution obtained through an empirical law calibrated using two fatigue experiments.
The microstructure enters the model by statistical representation in the RVE of the shape, size, and orientation distribution of the grains, which are generated from experimental data. The strategy allows fatigue life prediction for both strain- and stress-controlled experiments by using different FIPs, based on cyclic and total dissipated work, respectively. The framework has been applied to Hastelloy-X built in different longitudinal directions. The main conclusions about the modeling framework and its application to Hastelloy-X are the following.
\begin{itemize}
    \item A phenomenological crystal plasticity model that includes a back stress term can accurately reproduce the Hastelloy-X hysteresis curve under all loading conditions.
    \item Accelerating the evolution of internal variables so that the stable hysteresis cycle of the model is reached before experimental stabilization strongly reduces computational cost and provides an accurate result in fatigue prediction.
    \item A power-law life prediction model with two fitting parameters, based on cyclic and accumulated plastic work FIP, is able to accurately predict life for strain and stress controlled tests, respectively.
    \item The direction of the SLM fabrication (along the loading axis or perpendicular to it) {is found to influence} the fatigue performance of Hastelloy-X, especially for high applied stress ranges.
    \item The model can capture the effect of the fabrication direction on fatigue performance by using RVEs statistically representing the actual microstructure resulting from each fabrication direction.
    \item The model shows that grain elongation is the most relevant microstructural feature responsible for the better fatigue performance of Bulk-X samples.
\end{itemize}

 \bibliographystyle{elsarticle-num}





\end{document}